\documentclass[prc,twocolumn,preprintnumbers,showpacs,superscriptaddress]{revtex4}
\usepackage{amsmath,amssymb,epsfig,bm,color,slashed,
tikz,mathrsfs,graphicx,float,comment,isotope,mathrsfs}
\usepackage{cancel}
\usepackage{aas_macros}
\usepackage[normalem]{ulem}
\usepackage[colorlinks=true,linktocpage=true,linkcolor=blue,citecolor=magenta,urlcolor=blue]{hyperref}

\definecolor{red}{rgb}{0.8,0,0}
\definecolor{violet}{rgb}{0.4,0,0.4}
\definecolor{green}{rgb}{0,0.5,0.0}
\definecolor{navy}{rgb}{0.0,0.0,0.6}
\definecolor{orange}{rgb}{0.8,0.2,0.0}

\usepackage[normalem]{ulem}  

\newcommand{\bea}{\begin{eqnarray}}
\newcommand{\eea}{\end{eqnarray}}
\newcommand{\ep}{\epsilon}

\newcommand{\vecp}{{\bm p}}
\newcommand{\vecq}{{\bm q}}

\newcommand{\vecv}{{\bm v}}
\newcommand{\vecE}{{\bm E}}

\newcommand{\vecH}{{\bm H}}

\input{acronym.input}

\begin{document}

\title{Impact of positrons on electrical conductivity of hot and dense astrophysical plasma}

\author{Tigran Petrosyan}
\email{tigran.petrosyan.203@gmail.com }
\affiliation{Physics Institute, Yerevan State University, Yerevan 0025, Armenia}

\author{Arus Harutyunyan} \email{arus@bao.sci.am}
\affiliation{Byurakan Astrophysical Observatory,
  Byurakan 0213, Armenia}
\affiliation{Physics Institute, Yerevan State University, Yerevan 0025, Armenia}

\author{Armen Sedrakian}
\email{sedrakian@fias.uni-frankfurt.de}
\affiliation{Frankfurt Institute for Advanced Studies, D-60438
  Frankfurt am Main, Germany}
\affiliation{Institute of Theoretical Physics, University of Wroc\l{}aw,
50-204 Wroc\l{}aw, Poland}

\begin{abstract}
We study the influence of positrons on the outer crusts of neutron
stars and the interiors of white dwarfs, introducing them as a novel
component in both the composition of matter and in transport
processes. We solve a system of coupled Boltzmann kinetic equations
for the electron and positron distribution functions in the
relaxation-time approximation, taking into account electron–ion,
positron–ion, and electron–positron collisions. The relevant
scattering matrix elements are calculated from one-plasmon exchange
diagrams, with in-medium polarization tensors derived within the
hard–thermal–loop effective theory. Numerical results are obtained
for matter composed of carbon, iron and helium nuclei.  We find that the conductivity
rises with temperature, following a power law $\sigma \propto T^4$
in the semidegenerate regime and $\sigma \propto T$ in the
nondegenerate regime, due to the intense creation of thermal
electron–positron pairs and the resulting collisions among
them. These results highlight the importance of including positrons
in the transport properties of heated, dense astrophysical plasmas.
\end{abstract}

\maketitle

\section{Introduction}
\label{sec:intro}

Electrical conductivity of matter at subsaturation densities in
neutron stars and cores of white dwarfs plays a central role in several
astrophysical problems, such as the magnetic field
evolution~\cite{1992ApJ...395..250G,1995JApA...16..217B,1996ApJ...458..347M,Ascenzi2024},
dissipation of magnetohydrodynamic
waves~\cite{1979ApJ...227..995E,2014Ap.....57..530S}, etc.
The conductivity of dense neutron-star matter has been studied extensively
in the cold regime relevant to isolated stars
($T \leq 1$~MeV)~\cite{Flowers1976,Yakovlev1980,Urpin1980,Flowers1981,Itoh1983,Itoh1984,Nandkumar1984MNRAS,Sedrakian1987,Itoh1993,Baiko1998,Potekhin1999,Shternin2006,Itoh2008};
see Ref.~\cite{Schmitt2018} for a recent review.  However, formation
of hot crustal matter in compact stars is anticipated by various
astrophysical scenarios, such as supernova explosions leading to hot
protoneutron stars, remnants of binary neutron star mergers, as well
as heating of neutron star crusts in accreting binaries.  In this
regime, the plasma forms a liquid state of correlated ions, while the
relativistic electron gas becomes nondegenerate. This transition
triggers intense production of electron–positron pairs, introducing a
positron component into the plasma that contributes significantly to
its conductivity. Finite temperature properties of such a plasma are
expected to significantly deviate from zero‑temperature (cold,
degenerate) results.

The electrical conductivity of a one-component plasma in the hot
regime under magnetic fields was investigated in
Ref.~\cite{Harutyunyan2016}. These results were later employed in
Ref.~\cite{Harutyunyan2018} to estimate the impact of Ohmic
dissipation and the Hall effect on the dynamics of binary neutron star
mergers. In addition, the mean free path of conducting electrons was
used to identify the density–temperature regimes in which the
magnetohydrodynamic description breaks down.

The present work extends the analysis of Ref.~\cite{Harutyunyan2016}
to a hot plasma state in which a non-negligible population of
positrons coexists with electrons. Although positron contributions
were neglected in earlier studies, they become critically important in
the outer crust of neutron stars at temperatures
$T \gtrsim 1\textrm{MeV}=1.16\times 10^{10}$~K. In this regime,
electrons and positrons constitute the primary charge carriers, and
electrical conduction is governed mainly by electron–ion,
positron–ion, and electron–positron scattering processes.

To describe transport, we employ Boltzmann kinetic theory in the
relaxation-time approximation to obtain the nonequilibrium
distribution functions of electrons and positrons, fully accounting
for electron–ion, positron–ion, and electron–positron
collisions. In-medium screening effects are incorporated through the
structure factor of the one-component classical plasma and the
hard-thermal-loop effective-theory treatment of the plasmon
self-energy~\cite{Braaten1990,Braaten1992,Harutyunyan2016}.

This work is structured as follows. In Sec.~\ref{sec:Boltzmann} we
derive the electrical conductivity for a multicomponent plasma
consisting of electrons, positrons and ions, solving the system of
linearized coupled Boltzmann equations for electron and positron
distribution functions. In Sec.~\ref{sec:rates}, the matrix elements
for electron-ion, positron-ion, and positron-electron scattering
processes and the resulting collision rates are
discussed. Section~\ref{sec:results} presents the phase structure of
crustal matter and numerical results for electrical conductivity. Our
results are summarized in Sec.~\ref{sec:summary}.
Appendix~\ref{app:matrix} provides the details of the derivation of
the electron-positron scattering matrix element, and
Appendix~\ref{app:nu_ep} computes the collision rate due to this
scattering channel.

We use the natural units with $\hbar= c = k_B = k_e = 1$,
$e=\sqrt{\alpha}$, $\alpha=1/137$ and the metric signature
$(1,-1,-1,-1)$. Greek and Latin indices are utilized to label 4-space
and 3-space tensor quantities, respectively.

\section{Electrical conductivity from the Boltzmann equation}
\label{sec:Boltzmann}


The kinetics of electrons and positrons is described by a system of coupled Boltzmann equations 
\bea\label{eq:boltzmann-}
&&\frac{\partial f^-}{\partial t}+
\bm v\frac{\partial f^-}{\partial\bm r}-
e
(\bm E+\bm v \times \bm H)
\frac{\partial f^-}
{\partial\bm p}=I^-,\quad\\
\label{eq:boltzmann+}
&&\frac{\partial f^+}{\partial t}+
\bm v\frac{\partial f^+}{\partial\bm r}+
e
(\bm E+\bm v \times \bm H)
\frac{\partial f^+}
{\partial\bm p}=I^+,\quad
\eea
where $f^-$ and $f^+$ are the distribution functions for electrons and
positrons, respectively, $\vecE$ and $\vecH$ are the electric and
magnetic fields, $\vecv$ is the electron or positron velocity, and $I^-$
and $I^+$ are the collision integrals, respectively, for electrons and
positrons, which include electron-ion and electron-positron scattering
for $I^-$, and positron-ion and positron-electron scattering for $I^+$,
\bea\label{eq:coll_integral-}
I^-=\sum_i I_{ei}+I_{ep},\\
\label{eq:coll_integral+}
I^+=\sum_i I_{pi}+I_{pe}, 
\eea
where the sums in the first terms run over the ion species,
  allowing for the general case of a multicomponent plasma in which
  several types of ions may coexist. In general, the composition is
  described by a statistical distribution of nuclear species; however,
  in the present work, we assume that a single species is dominant,
  and we restrict our numerical computations to a plasma composed of
  this primary nuclear species. This approximation is valid in the
  low- to intermediate-temperature regime considered here, but may
  require corrections at higher temperatures, where differences in the
  binding energies of various nuclei become less significant in
  determining the composition of matter. A detailed study of a
  genuinely multicomponent plasma with temperature-dependent
  compositions is left for future work.  

For electron-ion and positron-ion collision integrals, we have
\bea\label{eq:collision_ei}
I_{ei} &=& -(2\pi)^4\sum\limits_{234}|{\cal M}_{12\to 34}^{ei}|^2
\delta^{(4)}(p_1+p_2-p_3-p_4)\nonumber\\
&&\hspace{-1cm}\times \Big[f^-_1g_2^i(1-f^-_3)(1\pm g_4^i)-f^-_3g_4^i(1-f^-_1)(1\pm g_2^i)\Big],\\
\label{eq:collision_pi}
I_{pi} &=& -(2\pi)^4\sum\limits_{234}|{\cal M}_{12\to 34}^{pi}|^2
\delta^{(4)}(p_1+p_2-p_3-p_4)\nonumber\\
&&\hspace{-1cm}\times\Big[f^+_1g_2^i(1-f^+_3)(1\pm g_4^i)-f^+_3g_4^i(1-f^+_1)(1\pm g_2^i)\Big],\quad
\eea
where the indices $1$ and $3$ label the incoming and outgoing
electrons or positrons, respectively, $g^i_{2,4} \equiv g^i(p_{2,4})$ are
the ion distribution functions before and after the collision, and
${\cal M}_{12\to 34}^{ei,pi}$ is the electron- or positron–ion scattering
matrix element. The upper signs in
Eqs.~\eqref{eq:collision_ei} and \eqref{eq:collision_pi} correspond to
Bose statistics and the lower signs to Fermi statistics for ions. We
also use the shorthand notation $\sum\limits_i = \int
d\vecp_i/(2\pi)^3$. Assuming that ions remain in thermal equilibrium,
we take $g^i(p)$ in the form of the Bose or Fermi distribution function,
$g^i(p) = \left[e^{\beta(\varepsilon_p - \mu_i)} \mp 1\right]^{-1},$
where $\varepsilon_p = p^2/M_i$, and $M_i$ and $\mu_i$ are the mass
and chemical potential of ions of type $i$, while $\beta = T^{-1}$ is the
inverse temperature. 

The electron-positron and positron-electron collision integrals can be written in a similar fashion
\bea\label{eq:collision_ep}
I_{ep} &=& -(2\pi)^4\sum\limits_{234}|{\cal M}_{12\to 34}^{ep}|^2
\delta^{(4)}(p_1+p_2-p_3-p_4)\nonumber\\
&&\hspace{-1.5cm}\times \Big[f^-_1f^+_2(1-f^-_3)(1-f^+_4)-f^-_3f^+_4(1-f^-_1)(1-f^+_2)\Big],\\
\label{eq:collision_pe}
I_{pe} &=& -(2\pi)^4\sum\limits_{234}|{\cal M}_{12\to 34}^{pe}|^2
\delta^{(4)}(p_1+p_2-p_3-p_4)\nonumber\\
&&\hspace{-1.5cm}\times \Big[f^+_1f^-_2(1-f^+_3)(1-f^-_4)-f^+_3f^-_4(1-f^+_1)(1-f^-_2)\Big],
\eea
where ${\cal M}_{12\to 34}^{ep}$ and ${\cal M}_{12\to 34}^{pe}$ are
the electron-positron and positron-electron scattering matrix
elements, respectively.

To solve the Boltzmann equations, we will consider weak deviations from equilibrium  
and search the distribution functions in the form
\bea\label{eq:distribution}
f^{\pm}= f^{0\pm}+\delta f^{\pm},\qquad \delta f^{\pm}=-\phi^{\pm}
\frac{\partial f^{0\pm}}{\partial\ep},
\eea
where
$
f^{0\pm}({\ep})=[e^{\beta(\ep \pm\mu_e)}+1]^{-1}
$
are the local Fermi distribution functions for positrons $(+)$ and
electrons $(-)$
with $\ep=\sqrt{p^2+m^2}$, $m$ is the electron mass, $\mu_e$ is the
electron chemical potential, $\delta f^{\pm}\ll f^{0\pm}$ are small
perturbations, and $\phi^{\pm}$ are unknown functions; additionally,
introducing fermionic velocity $\bm v =\partial\ep/\partial \bm p=\bm
p/\ep$, we give some of the derivatives that will be used below,
\bea\label{eq:fermi_deriv_p}
\frac{\partial f^{0\pm}}{\partial\bm p} &=&
\bm v\frac{\partial f^{0\pm}}{\partial\ep},
\quad\frac{\partial f^{0\pm}}{\partial\ep}=
-\beta f^{0\pm}(1-f^{0\pm}).\quad
\eea
We assume plasma is spatially uniform, time independent and external
fields are uniform and constant; i.e., there is no spatial variation
of the fermionic distribution function, with only momentum space
evolution. 

We next linearize the Boltzmann equation by substituting
Eq.~\eqref{eq:distribution} into Eqs.~\eqref{eq:boltzmann-} and
\eqref{eq:boltzmann+} and keeping only the terms linear in
$\delta f^{\pm}$ and in the electric field $\vecE$, which implies that
in the terms proportional to $\vecE$, the substitutions
$f^{\pm} \to f^{0\pm}$ can be made. We further neglect magnetic
fields, which induce anisotropy in the conductivity; for the
discussion of the effects of the $\vecH$ field on transport, see 
Ref.~\cite{Harutyunyan2016}. Thus, the linearized
Boltzmann equations take the form
\bea\label{eq:boltzmann_cond1-}
-\frac{\partial f^{0-}}{\partial \ep}
e\bm v\cdot\bm E= \sum_i I_{ei}+I_{ep},\\
\label{eq:boltzmann_cond1+}
\frac{\partial f^{0+}}{\partial \ep}
e\bm v\cdot\bm E= \sum_i I_{pi}+I_{pe},
\eea
where the linearized collision integrals are given by
\bea
\label{eq:collision_integrals-}
I_{ei}=-\sum_{ei} (\phi^-_1-\phi^-_3),~
I_{ep}= -\sum_{ep} (\phi^-_1+\phi^+_2-\phi^-_3-\phi^+_4),\nonumber\\
\label{eq:collision_integrals+}
I_{pi}=-\sum_{pi} (\phi^+_1-\phi^+_3),~
I_{pe}=-\sum_{pe} (\phi^+_1+\phi^-_2-\phi^+_3-\phi^-_4),\nonumber
\eea
where
\bea\label{eq:coll_ei_short}
\sum_{ei} &\equiv & (2\pi)^4\beta \sum\limits_{234}
|{\cal M}_{12\to 34}^{ei}|^2\delta^{(4)}(p_1+p_2-p_3-p_4)\nonumber\\
&\times&f^{0-}_1(1-f^{0-}_3)g_2^i(1\pm g_4^i),\\
\label{eq:coll_pi_short}
\sum_{pi} &\equiv & (2\pi)^4\beta \sum\limits_{234}
|{\cal M}_{12\to 34}^{pi}|^2
\delta^{(4)}(p_1+p_2-p_3-p_4)\nonumber\\
&\times&f^{0+}_1(1-f^{0+}_3)g_2^i(1\pm g_4^i),\\
\label{eq:coll_ep_short}
\sum_{ep} &\equiv & (2\pi)^4\beta\sum\limits_{234}|{\cal M}_{12\to 34}^{ep}|^2\delta^{(4)}(p_1+p_2-p_3-p_4)\nonumber\\
&\times&f_1^{0-}f_2^{0+}(1-f_3^{0-})(1-f_4^{0+}),\\
\label{eq:coll_pe_short}
\sum_{pe} &\equiv & (2\pi)^4\beta\sum\limits_{234}|{\cal M}_{12\to 34}^{pe}|^2\delta^{(4)}(p_1+p_2-p_3-p_4)\nonumber\\
&\times&f_1^{0+}f_2^{0-}(1-f_3^{0+})(1-f_4^{0-}).\qquad
\eea
We search for the solutions of Eqs.~\eqref{eq:boltzmann_cond1-} and
\eqref{eq:boltzmann_cond1+} in the form
\bea\label{eq:solution_cond5}
\phi^{\pm} = {\pm}e\tau^{\pm} (\bm v\cdot\bm E),
\eea
where the relaxation times $\tau^{\pm}$ depend on the particle energy
$\epsilon$. Substituting this into the collision integrals yields
\bea\label{eq:coll_ei1}
I_{ei} &=& e\bm E\sum_{ei}\Big(\tau^-_{1}\bm v_1-\tau^-_{3}\bm v_3\Big),\\
\label{eq:coll_pi1}
I_{pi} &=& -e\bm E\sum_{pi}\Big(\tau^+_{1}\bm v_1-\tau^+_{3}\bm v_3\Big),\\
\label{eq:coll_ep1}
I_{ep} &=& e\bm E\sum_{ep}\Big(\tau^-_{1}\bm v_1-\tau^+_{2}\bm v_2-\tau^-_{3}\bm v_3+\tau^+_{4}\bm v_4\Big),\\
\label{eq:coll_pe1}
I_{pe} &=& -e\bm E\sum_{pe}\Big(\tau^+_{1}\bm v_1-\tau^-_{2}\bm v_2-\tau^+_{3}\bm v_3+\tau^-_{4}\bm v_4\Big),\quad
\eea
and Eqs.~\eqref{eq:boltzmann_cond1-} and \eqref{eq:boltzmann_cond1+}
become (the cancellation of the vector $\bm E$ from both sides of
the equations is justified as these should be satisfied for an
arbitrary direction of $\bm E$; equivalently, one may project the
equation along $\bm E$ and divide by $\vert \bm E\vert$)
\bea\label{eq:boltzmann_cond6-}
-\frac{\partial f^{0-}}{\partial \ep}\bm v
&=& \sum_{i}\sum_{ei}\Big(\tau^-_{1}\bm v_1-\tau^-_{3}\bm v_3\Big)
\nonumber\\
&&\hspace{-1cm}+\sum_{ep}\Big(\tau^-_{1}\bm v_1-\tau^+_{2}\bm v_2-\tau^-_{3}\bm v_3+\tau^+_{4}\bm v_4\Big),\qquad\\
\label{eq:boltzmann_cond6+}
-\frac{\partial f^{0+}}{\partial \ep}\bm v
&=& \sum_{i}\sum_{pi}\Big(\tau^+_{1}\bm v_1-\tau^+_{3}\bm v_3\Big) \nonumber\\
&&\hspace{-1cm}+\sum_{pe}\Big(\tau^+_{1}\bm v_1-\tau^-_{2}\bm v_2-\tau^+_{3}\bm v_3+\tau^-_{4}\bm v_4\Big).\qquad
\eea
We multiply Eqs.~\eqref{eq:boltzmann_cond6-} and
\eqref{eq:boltzmann_cond6+} by
$\tau_1^- \vecv_1$ and $\tau_1^+ \vecv_1$, respectively,
and integrate over $\vecp_1 \equiv \vecp$,
yielding, 
\bea\label{eq:boltzmann_cond7-}
-\sum_1\frac{\partial f^{0-}}{\partial \ep}\tau^- \vecv^2
&=& \frac{1}{2}\sum_{1,i}\sum_{ei}\Big(\tau_1^-\bm v_1-\tau_3^- \bm v_3\Big)^2
\nonumber\\
&+&\frac{1}{2}\sum_1\sum_{ep}\Big(\tau_1^-\bm
v_1-\tau^-_3\bm v_3\Big) \nonumber\\
&&\hspace{-1.1cm}\times\Big(\tau^-_{1}\bm v_1-\tau^+_{2}\bm v_2-\tau^-_{3}\bm v_3+\tau^+_{4}\bm v_4\Big),\qquad\\
\label{eq:boltzmann_cond7+}
-\sum_1\frac{\partial f^{0+}}{\partial \ep}\tau^+ \vecv^2
&=& \frac{1}{2}\sum_{1,i}\sum_{pi}\Big(\tau_1^+\bm v_1-\tau_3^+ \bm v_3\Big)^2
\nonumber\\
&+&\frac{1}{2}\sum_1\sum_{pe}\Big(\tau_1^+\bm v_1-\tau^+_3\bm v_3\Big)
\nonumber\\
&&\hspace{-1.2cm}\times\Big(\tau^+_{1}\bm v_1-\tau^-_{2}\bm v_2-\tau^+_{3}\bm v_3+\tau^-_{4}\bm v_4\Big),\qquad
\eea
where we performed permutations $1\leftrightarrow 3$ and $2\leftrightarrow 4$.
To simplify the sums we make a simple ansatz $\tau^{\pm}(\ep)\propto \ep$, which gives ($\ep\equiv\ep_1$)
\bea\label{eq:sum_ei1}
\sum_{e,pi}\Big(\tau^\pm_{1}\bm v_1
-\tau^\pm_{3}\bm v_3\Big)^2
=\left(X^\pm\right)^2\sum_{e,pi} \vecq^2,
\eea
where $\bm q =\bm p_1-\bm p_3 = \bm p_4-\bm p_2$ is the transferred
momentum, and $X^\pm={\tau^\pm}(\ep)/{\ep}={\rm const}$. Similarly, 
\bea\label{eq:sum_ep1}
&&\sum_{ep}\Big(\tau_1^-\bm v_1-\tau^-_3\bm v_3\Big)
\Big(\tau^-_{1}\bm v_1-\tau^+_{2}\bm v_2-\tau^-_{3}\bm v_3+\tau^+_{4}\bm v_4\Big)\nonumber\\
&& \hspace{2cm}= X^-\left(X^- + X^+\right)\sum_{ep}\bm q^2,\\
\label{eq:sum_pe1}
&&\sum_{pe}\Big(\tau_1^+\bm v_1-\tau^+_3\bm v_3\Big)
\Big(\tau^+_{1}\bm v_1-\tau^-_{2}\bm v_2-\tau^+_{3}\bm
v_3+\tau^-_{4}\bm v_4\Big) \nonumber\\
&&\hspace{2cm} = X^+\left(X^- + X^+\right)\sum_{pe}\bm q^2.
\eea
Then Eqs.~\eqref{eq:boltzmann_cond7-} and \eqref{eq:boltzmann_cond7+}
can be written as 
\bea\label{eq:boltzmann_cond11-}
X^{-}\bigg(\sum_{i}\nu_{ei}+\nu_{ep}\bigg)+X^{+}\nu_{ep}=1,\\
\label{eq:boltzmann_cond11+}
X^{+}\bigg(\sum_{i}\nu_{pi}+\nu_{pe}\bigg)+X^{-}\nu_{pe}=1,
\eea
with
\bea\label{eq:nu_ei_pi}
\nu_{ei}&=&\bigg[-\sum_1\frac{\partial f^{0-}}{\partial \epsilon}
\frac{p^2}{\ep}\bigg]^{-1}\frac{1}{2}\sum_1\sum_{ei} \vecq^2,\\
\label{eq:nu_ep_pe}
\nu_{ep}&=&\bigg[-\sum_1\frac{\partial f^{0-}}{\partial \ep}
\frac{p^2}{\ep}\bigg]^{-1}\frac{1}{2}\sum_1\sum_{ep} \vecq^2.
\eea
Similar expressions hold for positrons with the replacements
$ei\to pi$, $ep\to pe$ and $f^{0-} \to f^{0+}$.  These quantities
represent appropriately weighted {\it mean collision rates} between
the corresponding particles. Solving Eqs.~\eqref{eq:boltzmann_cond11-}
and \eqref{eq:boltzmann_cond11+} then yields
\bea\label{eq:X-}
X^{-} &=& \frac{\sum_i\nu_{pi}+\nu_{pe}-\nu_{ep}}{\big(\sum_{i}\nu_{ei}+\nu_{ep}\big)\big(\sum_{i}\nu_{pi}+\nu_{pe}\big)-\nu_{ep}\nu_{pe}},\\
\label{eq:X+}
X^{+} &=& \frac{\sum_i\nu_{ei}+\nu_{ep}-\nu_{pe}}{\big(\sum_{i}\nu_{ei}+\nu_{ep}\big)\big(\sum_{i}\nu_{pi}+\nu_{pe}\big)-\nu_{ep}\nu_{pe}}.\quad
\eea
In the case where electron-positron collisions are neglected, i.e.,
$\nu_{ep}=\nu_{pe}=0$, Eqs.~\eqref{eq:X-} and \eqref{eq:X+} reduce to
the well-known expressions
\bea\label{eq:tau_effective_ei}
(X^{-})^{-1} =\sum_i \nu_{ei},\qquad
(X^{+})^{-1} =\sum_i \nu_{pi},
\eea
where the collision rates on different types of nuclei are summed up.

In the regimes where electron-positron collisions become important,
the numbers of electrons and positrons are almost equal, and we find
that, practically always, $\nu_{ep}\simeq\nu_{pe}$, such that
$\nu_{ep}-\nu_{pe}\ll \nu_{ei}\simeq\nu_{pi}$. Then, Eqs.~\eqref{eq:X-}
and \eqref{eq:X+} can be simplified to
\bea\label{eq:X-1}
X^{-} &=& \frac{\sum_i\nu_{pi}}{\big(\sum_{i}\nu_{ei}\big)\big(\sum_{i}\nu_{pi}\big)+\nu_{ep}\sum_{i}(\nu_{ei}+\nu_{pi})},\quad\\
\label{eq:X+1}
X^{+} &=& \frac{\sum_i\nu_{ei}}{\big(\sum_{i}\nu_{ei}\big)\big(\sum_{i}\nu_{pi}\big)+\nu_{ep}\sum_{i}(\nu_{ei}+\nu_{pi})}.\quad
\eea
In the limit of fast electron-positron scattering we have also
$\nu_{ei}\simeq\nu_{pi}\ll \nu_{ep}\simeq\nu_{pe}$; therefore,
\bea\label{eq:tau_effective_ep}
X^{-}=X^{+}=\frac{1}{2\nu_{ep}}=\frac{1}{2\nu_{pe}},
\eea
which can also be derived directly from
Eqs.~\eqref{eq:boltzmann_cond11-} and \eqref{eq:boltzmann_cond11+}.

We next write explicit expressions for the mean collision rates by
substituting Eqs.~\eqref{eq:coll_ei_short}–\eqref{eq:coll_pe_short}
into \eqref{eq:nu_ei_pi} and \eqref{eq:nu_ep_pe} (and their analogs
for positrons) to obtain
\bea\label{eq:nu_ei1}
\nu_{ei} &=& \frac{(2\pi)^4}{2l^-} \!\sum\limits_{1234}
|{\cal M}_{12\to 34}^{ei}|^2
\delta^{(4)}(p_1+p_2-p_3-p_4)\nonumber\\
&& \times f^{0-}_1(1-f^{0-}_3)g_2^i(1\pm g_4^i) \vecq^2,\\
\label{eq:nu_pi1}
\nu_{pi} &=& \frac{(2\pi)^4}{2l^+} \!\sum\limits_{1234}
|{\cal M}_{12\to 34}^{pi}|^2
\delta^{(4)}(p_1+p_2-p_3-p_4)\nonumber\\
&& \times  f^{0+}_1(1-f^{0+}_3)g_2^i(1\pm g_4^i) \vecq^2,\\
\label{eq:nu_ep1}
\nu_{ep} &=& \frac{(2\pi)^4}{2l^-} 
\sum\limits_{1234}
|{\cal M}_{12\to 34}^{ep}|^2\delta^{(4)}(p_1+p_2-p_3-p_4) \nonumber\\
&&\times  f_1^{0-}f_2^{0+}(1-f_3^{0-})(1-f_4^{0+}) \vecq^2,\\
\label{eq:nu_pe1}
\nu_{pe} &=& \frac{(2\pi)^4}{2l^+} 
\sum\limits_{1234}
|{\cal M}_{12\to 34}^{pe}|^2\delta^{(4)}(p_1+p_2-p_3-p_4)\nonumber\\
&&\times f_1^{0+}f_2^{0-}(1-f_3^{0+})(1-f_4^{0-}) \vecq^2,
\eea
with
\bea\label{eq:l_pm}
l^{\pm}\equiv\frac{1}{2\pi^2}\int_m^\infty\! d\ep\, p^3 f^{0\pm}\left(1-f^{0\pm}\right),
\eea
where we used Eq.~\eqref{eq:fermi_deriv_p}.

We can now calculate the contributions of electrons and positrons to the electric current:
\bea\label{eq:current_e}
j^-_k &=& -\int\!\frac{2d\bm p}{(2\pi)^3}ev_k\,\delta f^{-}
= \sigma^-_{kj}E_j,\\
\label{eq:current_p}
j^+_k &=& \int\!\frac{2d\bm p}{(2\pi)^3}ev_k\,\delta f^{+} 
= \sigma^+_{kj}E_j,
\eea
where we used Eqs.~\eqref{eq:distribution} and
\eqref{eq:solution_cond5} and defined the conductivity matrices (the
factor 2 accounts for spin degeneracy) as
\bea\label{sigma_kj}
\sigma^{\pm}_{kj} = -\int\!\frac{2d\bm p}{(2\pi)^3}\frac{\partial f^{0\pm}}{\partial\ep} e^2\tau^{\pm}v_kv_j
=\delta_{kj}\sigma^{\pm},
\eea
where isotropy of the medium allows us to define the scalar conductivities:
\bea\label{eq:sigma}
\sigma^{\pm}=-\frac{e^2}{3\pi^2}\int_m^\infty d\ep\,\frac{p^3}{\ep}
\frac{\partial f^{0\pm}}{\partial\ep}\tau^{\pm}.
\eea
The total current is then
\bea\label{eq:current_sum}
\bm j=\bm j^++\bm j^- = {\sigma} \bm E,
\eea
with total conductivity
${\sigma}={\sigma}^++{\sigma}^-$. The electrical conductivity is thus
fully determined once the relaxation times $\tau^{\pm}$ are known.
By employing the ansatz $\tau^\pm=\ep X^\pm$, 
Eq.~\eqref{eq:sigma} becomes
\bea\label{eq:sigma1}
\sigma^{\pm}=\frac{e^2 X^{\pm}}{3\pi^2 T}\!\int_m^\infty d\ep\,{p^3}f^{0\pm}\left(1-f^{0\pm}\right) = \frac{2e^2}{3T} X^{\pm} l^{\pm}.
\eea

\begin{figure*}[t]
\includegraphics[width=12cm, keepaspectratio]{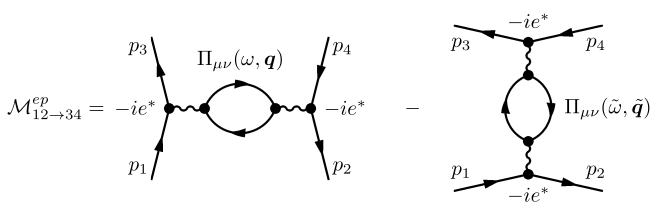}
\caption{Diagrams describing the electron-positron scattering and
  annihilation via exchange of a virtual plasmon. The plasmon
  self-energy is given by the polarization tensor
  $\Pi_{\mu\nu}(\omega,\vecq)$ shown by the closed loops.  }
\label{fig:ep_scattering}
\end{figure*}
 We remark that the range of validity of the Boltzmann kinetic
  theory employed here can be derived from the common
  considerations~\cite{Lifshitz_Pitaevskii1981}.  The main assumption
  is that the collisions are instantaneous, which is justified if the
  average scattering length which has the order of the Debye screening
  radius $r_D=q_D^{-1}$ of the electric field (see
  Sec.~\ref{sec:ei_matrix} for details) is much smaller than the
  electron or positron mean free paths $l^{\pm}\simeq \tau^{\pm}$
  between collisions (assuming ultra-relativistic particles with
  average velocities $\bar{v}^{\pm}\simeq 1$). The particle mean
  collision times and related mean free paths are supposed to be much
  smaller than the characteristic length scale $L$ and the inverse
  frequency $\Omega^{-1}$ of macroscopic perturbations of the electric
  field. Thus, the general validity condition can be written as
\bea 
q_D^{-1}\ll \tau^{\pm} \ll {\rm min}\{L,\Omega^{-1}\}.\nonumber
\eea
Our numerical results (see Sec.~\ref{sec:results},
Figs.~\ref{fig:mu_qd}--\ref{fig:tau_temp}) show that in the regime of
interest $0.03\leq q_D\leq 2$~MeV and
$10^2\leq\tau^{\pm}\leq 10^4$~MeV$^{-1}$ (recall that
1\,s$\,=\,1.52\times 10^{21}$~MeV$^{-1}$); therefore the condition
$q_D\tau^{\pm}\gg 1$ is satisfied practically in the whole regime of
interest with high accuracy.

\section{Scattering amplitudes and collision rates}
\label{sec:rates}

In this section, we will discuss the scattering matrix elements which
enter the integrands of collision
rates~\eqref{eq:nu_ei1}--\eqref{eq:nu_pe1}, using the standard QED
techniques for a thermal medium. In addition, we will also discuss the
electron-positron collision rates using the obtained expressions for
the matrix elements.

\subsection{Electron-positron scattering matrix element}
\label{sec:ep_matrix}

The screened electron–positron scattering matrix element can be
expressed as the sum of two contributions, corresponding to proper
scattering and annihilation processes (analogous to Bhabha scattering
in vacuum); see Fig.~\ref{fig:ep_scattering}.
\bea\label{eq:amplitude_ep}    
{\cal M}_{12\to 34}^{ep} &=&
{\cal M}_{12\to 34}^{ep,t}-
{\cal M}_{12\to 34}^{ep,s},
\eea
where each of the channels in a thermal medium can be split into
longitudinal and transverse parts (see Ref.~\cite{Harutyunyan2016} and
references therein)
\bea\label{eq:amplitude_ep_t}
{\cal M}_{12\to 34}^{ep,t} =
-{\cal M}^t_L+{\cal M}^t_T,~  {\cal M}^t_L = \frac{J_0J'_0}{t_0},~
{\cal M}^t_T = \frac{\bm J_\perp\bm J'_\perp}{t_\perp},\nonumber\\\\
\label{eq:amplitude_ep_s}
{\cal M}_{12\to 34}^{ep,s}
= -{\cal M}^s_L+{\cal M}^s_T,~
{\cal M}^s_L = \frac{\tilde{J}_0\tilde{J}'_0}{s_0},~
{\cal M}^s_T = \frac{\tilde{\bm J}_\perp\tilde{\bm J}'_\perp}{s_\perp},\nonumber\\
\eea
where we introduced the following (Mandelstam) variables
\bea\label{eq:t_s_channels}
t_0=q^2+\Pi_L,\quad 
t_\perp = q^2-\omega^2+\Pi_T,\\
s_0=\tilde{q}^2+\tilde{\Pi}_L,\quad 
s_\perp =\tilde{q}^2-\tilde{\omega}^2
+\tilde{\Pi}_T,
\eea
and 4-currents
\bea\label{eq:currents_ep}
J^{\mu}=-e^*\bar{u}_3 \gamma^\mu u_1,\quad
J'^{\mu}=-e^*\bar{v}_2 \gamma^\mu v_4,\\
\label{eq:currents_tilde_ep}
\tilde{J}^{\mu}=-e^*\bar{v}_2 \gamma^\mu u_1,\quad
\tilde{J}'^{\mu}=-e^*\bar{u}_3 \gamma^\mu v_4.
\eea
Here, $e^* = \sqrt{4\pi}e$, $(\omega,\vecq)=p_1-p_3=p_4-p_2$,
$(\tilde{\omega},\tilde{\vecq})=p_1+p_2=p_3+p_4$,
$u_a = {u}^{s_a}(p_a)$, $v_b = {v}^{s_b}(p_b)$,
$\bm J_\perp, \bm J'_\perp$ are the components of these currents
transverse to $\bm q$, and $\tilde{\bm J}_\perp, \tilde{\bm J}'_\perp$
are those transverse to $\tilde{\bm q}$.  The screening of the
interaction is taken into account in terms of the longitudinal
$\Pi_L(\omega,q) $ and transverse $\Pi_T(\omega, q)$ components of the
polarization tensor.

\begin{figure*}[t]
\includegraphics[width=8cm, keepaspectratio]{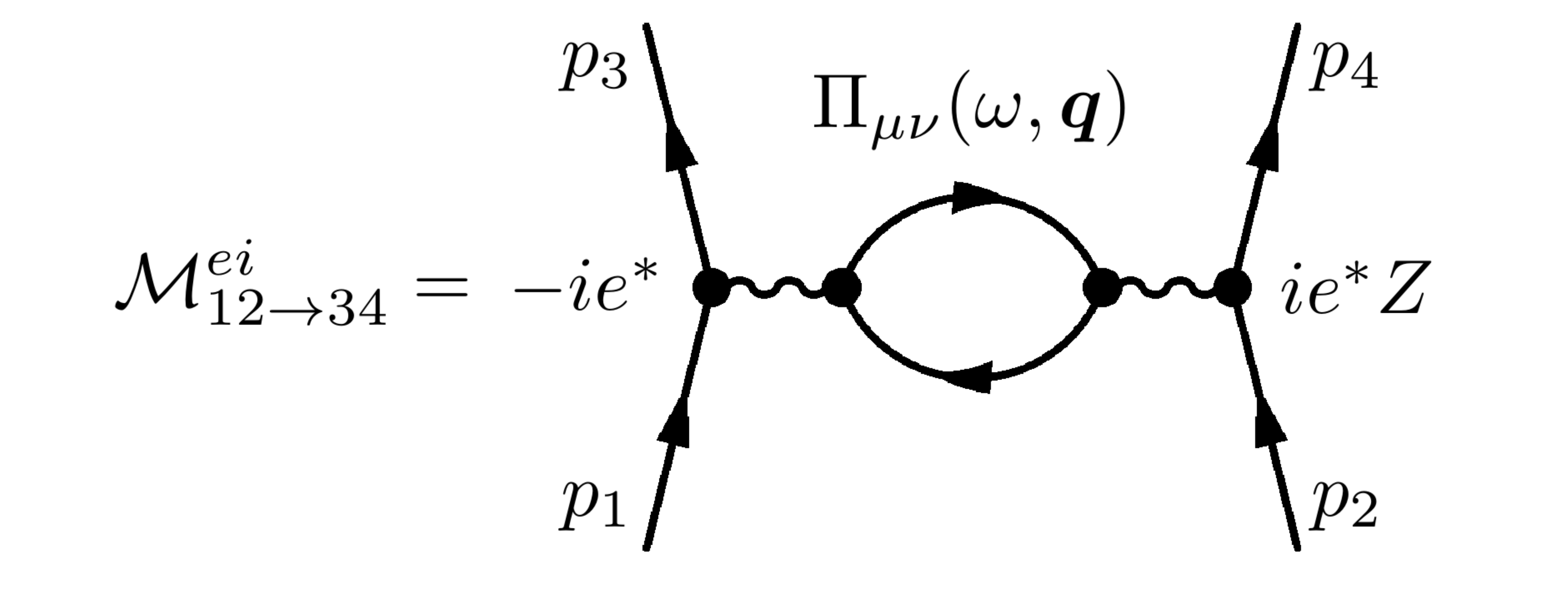}
\hspace{0.5cm}
\includegraphics[width=7cm, keepaspectratio]{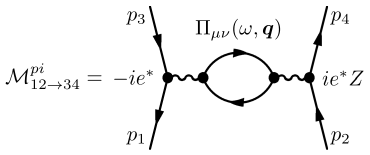}
\caption[] {Diagrams describing the electron-ion (left) and positron-ion (right) scattering via exchange of a virtual plasmon.
}
\label{fig:ei_scattering}
\end{figure*}

Details of the squaring and spin averaging of the $ep$ matrix element
are given in Appendix~\ref{app:matrix}. In this work, for simplicity,
we neglect  the
interference terms between the $t$- and $s$-channels in the
electron–positron scattering amplitude. The final results for the $t$-
and $s$-channel spin-averaged scattering amplitudes are 
\bea\label{eq:probability_ep4_t}
\frac{1}{4}\sum\limits_{\rm spins}|{\cal M}^{ep,t}_{12\to
  34}|^2&=&
\frac{e^{*4}}{4\ep_1\ep_2(\ep_1-\omega)(\ep_2+\omega)}\nonumber\\
&&\hspace{-2cm}\times\Bigg[\frac{C_0(\alpha,\theta)}{|t_0|^2}
+ \frac{C_2(\alpha,\theta)+4C^2_1(\alpha,\theta)\cos^2\varphi}{|t_\perp|^2}
\nonumber\\
&&\hspace{-2cm} -2{\rm Re}\frac{(2\ep_1-\omega)(2\ep_2+\omega)C_1(\alpha,\theta)\cos\varphi}{(t_0t^*_\perp)}
\Bigg],\\
\label{eq:probability_ep4_s}
\frac{1}{4}\sum\limits_{\rm spins}|{\cal M}^{ep,s}_{12\to
  34}|^2&=&
\frac{e^{*4}}{4\ep_1\ep_3(\tilde{\omega}-\ep_1)(\tilde{\omega}-\ep_3)}\nonumber\\
&&\hspace{-2cm}\times\Bigg[\frac{\tilde{C}_0(\alpha,\theta)}{|s_0|^2}
+\frac{\tilde{C}_2(\alpha,\theta)+4\tilde{C}^2_1(\alpha,\theta)\cos^2\varphi}{|s_\perp|^2}\nonumber\\
&&\hspace{-2cm} -2{\rm Re}\frac{(2\ep_1-\tilde{\omega})(2\ep_3-\tilde{\omega})\tilde{C}_1(\alpha,\theta)\cos\varphi}
{(s_0s^*_\perp)}\Bigg],\qquad
\eea
where, using the shorthands $\cos\alpha = x_\alpha$ and $\cos\theta =
x_\theta$,
\bea\label{eq:C0}
C_0&=& 
(2\epsilon_1^2 - \epsilon_1 \omega - p_1 q x_\alpha)
(2\epsilon_2^2 + \epsilon_2 \omega + p_2 q x_\theta),\\
\label{eq:C1}
C_1 &=& p_1 p_2 (1-x_\alpha^2)^{1/2}(1-x_\theta^2)^{1/2},\\
\label{eq:C2}
C_2 &=&
2 p_1^2 (1-x_\alpha^2) (\epsilon_2 \omega - p_2 q x_\theta) \nonumber\\
&+& 2 p_2^2 (1-x_\theta^2) (p_1 q x_\alpha - \epsilon_1 \omega) \nonumber\\
&+& 2 (p_1 q x_\alpha - \epsilon_1 \omega)(\epsilon_2 \omega - p_2 q x_\theta),\\
\label{eq:C0_tilde}
\tilde{C}_0 &=& 
(2\epsilon_1^2 - \epsilon_1 \tilde{\omega} - p_1 \tilde{q} x_\alpha)
(2\epsilon_3^2 - \epsilon_3 \tilde{\omega} - p_3 \tilde{q} x_\theta),\\
\label{eq:C1_tilde}
\tilde{C}_1 &=& p_1 p_3 (1-x_\alpha^2)^{1/2}(1-x_\theta^2)^{1/2},\\
\label{eq:C2_tilde}
\tilde{C}_2&=&
2 p_1^2 (1-x_\alpha^2) (p_3 \tilde{q} x_\theta - \epsilon_3 \tilde{\omega}) \nonumber\\
&+& 2 p_3^2 (1-x_\theta^2) (p_1 \tilde{q} x_\alpha - \epsilon_1 \tilde{\omega}) \nonumber\\
&+& 2 (\epsilon_1 \tilde{\omega} - p_1 \tilde{q} x_\alpha)(\epsilon_3 \tilde{\omega} - p_3 \tilde{q} x_\theta).
\eea

\subsection{Electron-ion and positron-ion scattering matrix elements}
\label{sec:ei_matrix}

In analogy to Eq.~\eqref{eq:amplitude_ep}, the one-plasmon-exchange
scattering amplitudes of electrons or positrons off ions can be written as (see Fig. \ref{fig:ei_scattering}) 
\bea\label{eq:amplitude_ei}
{\cal M}_{12\to 34}^{ei}&=&-\frac{J_0\bar{J}_0}{t_0}+
\frac{\bm J_\perp\cdot \bar{\bm J}_\perp}{t_\perp},\\
\label{eq:amplitude_pi}
{\cal M}_{12\to 34}^{pi}&=&-\frac{J'_0\bar{J}_0}{t_0}+
\frac{\bm J'_\perp\cdot \bar{\bm J}_\perp}{t_\perp},
\eea
where the electron, positron, and ion 4-currents are given, respectively, by
\bea\label{eq:currents}
&&J^{\mu}=-e^*\bar{u}_3\gamma^\mu u_1,\quad
J'^{\mu}=-e^*\bar{v}_1\gamma^\mu v_3,\\
&&\bar{J}^{\mu}
=Z_ie^*(1,\bm p_{24}/M_i),
\eea
and $\bm J_\perp, \bm J'_\perp$ are the components of the currents 
transverse to $\bm q$. Here, $\bm p_{24}=(\bm p_2+\bm p_4)/2$;
therefore, as $\bm p_4=\bm p_2+\bm q$, we obtain 
$\bm p_{24}^\perp =\bm p_{2}^\perp=\bm p_{4}^\perp\equiv \bm p'^{\perp}$.
Squaring these matrix elements and performing the spin-averaging
procedure yields the final result (see Ref.~\cite{Harutyunyan2016}),
\bea\label{eq:probability3}
\frac{1}{2}\sum\limits_{\rm spins}|{\cal M}^{p,ei}_{12\to 34}|^2&=&
 \frac{Z_i^2e^{*4}}{2\ep_1(\ep_1-\omega)}\nonumber\\
 &&\hspace{-2.5cm}\times \Biggl[
 \frac{\bar{C}_0(\alpha)}{|t_0|^2}-{\rm Re} 
\frac{2(2\ep_1-\omega) C_1(\alpha,\theta)\cos\varphi}{M_i (t_0
  t_\perp^*)}\nonumber\\
&&\hspace{-1.cm}+\frac{\bar{C}_2(\alpha,\theta)  +2C^2_1(\alpha,\theta)\cos^2\varphi}{M^2_i|t_\perp|^2}\Biggr],\quad
\eea
with
\bea\label{eq:C0_bar}
\bar{C}_0(\alpha) &=& 2 \epsilon_1^2 - \epsilon_1 \omega - p_1 q \cos \alpha,\\
\label{eq:C2_bar}
\bar{C}_2(\alpha,\theta) &=& p_2^2 \sin^2\theta \, (p_1 q \cos \alpha - \epsilon_1 \omega).
\eea

To account for ion–ion correlations and the finite size of the nuclei,
the $ei$ and $pi$ matrix elements are multiplied by $S_i(q) F_i^2(q)$,
where $S_i(q)$ is the ionic static structure factor and $F_i(q)$ is
the nuclear form factor~\cite{1984ApJ...285..758I}. (The effect of
$F_i(q)$ on transport in the outer crust is, however, small.) Our
numerical calculations are performed for a plasma consisting of a
single ion species, using the fit formulas for the static structure
factor $S(q)$ from Ref.~\cite{Desbiens2016}.

For the polarization tensor, we use the hard-thermal-loop expressions
of QED; see Ref.~\cite{Harutyunyan2016}, Eqs.~(43) and (44).  This
approximation assumes that $\omega\ll\ep$ and $q\ll p$; i.e., the
4-momentum of the plasmon is much smaller than that of the electron or
positron inside the fermionic loop. Numerical tests show that in the
whole range of densities and temperatures studied here, we effectively
have $\omega\lesssim q\ll p\simeq \ep$; therefore, the
hard-thermal-loop approximation to the polarization tensor works very
well.  In the regime of low-frequency scattering
$(x = \omega/q \ll 1)$, relevant for electron-ion and positron-ion
collisions, the longitudinal and transverse components (assuming
ultrarelativistic electrons and positrons) reduce to
\begin{align} \label{eq:Pi_lt_limit}
\Pi_L(q,\omega) = q_D^2  \chi_l,\quad
\Pi_T(q,\omega) = q_D^2  \chi_t,
\end{align}
with
${\rm Re}\chi_l = 1-x^2$,  ${\rm Im}\chi_l = -{\pi x}/{2}$,
${\rm Re}\chi_t = x^2$, and  ${\rm Im}\chi_t = {\pi x}/{4}.$
Here, $q_D$ is the Debye wave number, given by
\bea\label{eq:Debye}
q_D^2 = -\frac{4 e^2}{\pi} \int_m^\infty d\epsilon \, p  \,\epsilon \left( \frac{\partial f^{0+}}{\partial \epsilon} + \frac{\partial f^{0-}}{\partial \epsilon} \right).
\eea

\subsection{Collision rates}
\label{sec:nu_rates}

In this section, we summarize the results of the computations of the
collision rates given by
Eqs.~\eqref{eq:nu_ei1}--\eqref{eq:nu_pe1}. The electron-ion collision
rate was computed in Ref.~\cite{Harutyunyan2016} with the assumption
that the ion distribution is given in the classical Maxwell-Boltzmann
form.  This result can be generalized straightforwardly to obtain the
electron-ion and positron-ion collision rates~\eqref{eq:nu_ei1} and
\eqref{eq:nu_pi1}.  We will use the simplified formula given by
Eq.~(32) of Ref.~\cite{Harutyunyan2016}, which corresponds to the
limit of static scattering (this approximation is well justified for
ions with $Z_i>1$). The final formulas read
\bea\label{eq:nu_sigma_ei_final}
\nu_{pi/ei} &=& \frac{e^4 Z_i^2n_i}{2\pi l^{\pm}}\!
\int_{m}^\infty \!d\ep \,
f^{0\pm}(\ep)\left[1-f^{0\pm}(\ep)\right]\nonumber\\
&&\times\int_{0}^{2p} dq\, q^3S_i(q)F_i^2(q)\frac{4\ep^2-q^2}{|q^2+\Pi_L|^2}.\quad
\eea
The electron-positron collision rate is computed in Appendix~\ref{app:nu_ep}; the final result reads
\bea\label{eq:nu_ep_final}
\nu_{ep} &=& 
\frac{e^{4}}{(2\pi)^{3}l^-}\!\int_m^\infty\! d\ep\! \int_m^\infty\! d\ep'\!
\int_{m-\ep'}^{\ep-m} d\omega\,
\nonumber\\
&\times&f^{0-}(\ep)f^{0+}(\ep')\left[1-f^{0-}(\ep-\omega)\right]\left[1-f^{0+}(\ep+\omega)\right]\nonumber\\
&\times & \int_{Q_-}^{Q_+} dq\, \Bigg\{\frac{N_L}{4|q^2+\Pi_L|^2}+\frac{N_T}{2|q^2-\omega^2+\Pi_T|^2}\Bigg\}q^2\nonumber\\
&+&\frac{e^{4}}{(2\pi)^{3}l^-}\!\int_m^\infty\! d\ep\! \int_m^\infty\! d\ep'\!
\int_{\omega_{\rm min}}^{\infty} d{\omega}\nonumber\\
&\times&f^{0-}(\ep)f^{0+}({\omega}-\ep)\left[1-f^{0-}(\ep')\right]\left[1-f^{0+}({\omega}-\ep')\right]\nonumber\\
&\times & \int_{\tilde{Q}_-}^{\tilde{Q}_+} d{q}\, \Bigg\{\frac{u\tilde{N}_L}{4|{q}^2+{\Pi}_L|^2}+\frac{u\tilde{N}_T}{2|{q}^2-{\omega}^2
  +{\Pi}_T|^2}\nonumber\\
&+&{\rm Re}\frac{\tilde{N}_{LT}}{({q}^2+{\Pi}_L)({q}^2-{\omega}^2+{\Pi}_T)^*}\Bigg\},\qquad
\eea
where the first and second terms represent the scattering and annihilation diagrams, respectively.
Note that the variable $\ep$ stands for the energy of the initial state electron in both integrals, whereas the variable $\ep'$ represents the energy of the initial positron in the scattering integral and that of the final electron in the annihilation integral. Thus, the variables $q$ and $\omega$ are the momentum and energy of the transferred plasmon in both cases, but the integration limits are different.
The auxiliary functions in Eq.~\eqref{eq:nu_ep_final} are given by
\bea\label{eq:NL}
N_L &=& \big[(2\ep-\omega)^2-q^2\big]\big[(2\ep'+\omega)^2-q^2\big],\\
\label{eq:NT}
N_T &=&
\big[2p^2(1-x^2)+q^2-\omega^2\big]\big[2p'^2(1-y^2)+q^2-\omega^2\big],
\nonumber\\\\
\label{eq:NL_tilde}
\tilde{N}_L &=& \big[(2\ep-\omega)^2-q^2\big]\big[(2\ep'-\omega)^2-q^2\big],\\
\label{eq:NT_tilde}
\tilde{N}_T &=& \big[2p^2(1-{x}^2)+{q}^2-{\omega}^2\big]\big[2p'^2(1-z^2)+{q}^2-{\omega}^2\big], \nonumber\\\\
\label{eq:NLT_tilde}
\tilde{N}_{LT} &=&
2p^2p'^2(2\ep-{\omega})(2\ep'-{\omega})(1-{x}^2)(1-z^2),
\eea
where
\bea
\label{eq:xyz_main}
x&=&\frac{q^2-\omega^2+2\ep\omega}{2pq}, \quad y=\frac{\omega^2-q^2+2\ep'\omega}{2p'q},\\
z&=&\frac{{q}^2-{\omega}^2+2\ep'{\omega}}{2p'{q}},\quad u = p^2+p'^2-2pp'xz.\quad
\eea
The limits of the $q$-integration are given by $Q_+={\rm min}(q_+, q'_+)$, $Q_-={\rm max}(q_-, q'_-)$ with
\bea
q_{\pm} &=& \left\vert \sqrt{(\ep-\omega)^2-m^2} \pm \sqrt{\ep^2-m^2}\right\vert,
\\ q'_{\pm} &=& \left\vert  \sqrt{(\ep'+\omega)^2-m^2} \pm \sqrt{\ep'^2-m^2}\right\vert,
\eea
for the scattering integral. Similarly, for the annihilation integral we have $\tilde{Q}_+={\rm min}(\tilde{q}_+, \tilde{q}'_+)$, $\tilde{Q}_-={\rm max}(\tilde{q}_-, \tilde{q}'_-)$, with
\bea
\tilde{q}_{\pm} = {q}_{\pm},
\quad \tilde{q}'_{\pm} = \left\vert  \sqrt{(\ep'-{\omega})^2-m^2} \pm \sqrt{\ep'^2-m^2}\right\vert.
\eea
The lower bound for $\omega$ in the annihilation integral is
$\omega_{\rm min} = {\rm max}\{\ep,\ep'\}+m$ since the plasmon
produced in the annihilation carries the total energy of the
electron–positron pair.

Because the matrix element for $pe$ collisions is the same as the one
for $ep$ collisions (see Appendix~\ref{app:matrix}), we see from
Eqs.~\eqref{eq:nu_ep1} and \eqref{eq:nu_pe1} that the $pe$ collision
rate can be obtained from the $ep$ collision rate by simply replacing
$f^{0+}\leftrightarrow f^{0-}$ and $l^-\rightarrow l^+$; therefore, 
\bea\label{eq:nu_pe_final}
\nu_{pe} &=& 
\frac{e^{4}}{(2\pi)^{3}l^+}\!\int_m^\infty\! d\ep\! \int_m^\infty\! d\ep'\!
\int_{m-\ep'}^{\ep-m} d\omega\,f^{0+}(\ep)f^{0-}(\ep')\nonumber\\
&&\left[1-f^{0+}(\ep-\omega)\right]\left[1-f^{0-}(\ep+\omega)\right]\nonumber\\
&\times & \int_{Q_-}^{Q_+} dq\, \Bigg\{\frac{N_L}{4|q^2+\Pi_L|^2}+\frac{N_T}{2|q^2-\omega^2+\Pi_T|^2}\Bigg\}q^2\nonumber\\
&+&\frac{e^{4}}{(2\pi)^{3}l^+}\!\int_m^\infty\! d\ep\! \int_m^\infty\! d\ep'\!
\int_{\omega_{\rm min}}^{\infty} d{\omega}\nonumber\\
&&f^{0+}(\ep)f^{0-}({\omega}-\ep)\left[1-f^{0+}(\ep')\right]\left[1-f^{0-}({\omega}-\ep')\right]\nonumber\\
&\times & \int_{\tilde{Q}_-}^{\tilde{Q}_+} d{q}\, \Bigg\{\frac{u\tilde{N}_L}{4|{q}^2+{\Pi}_L|^2}+\frac{u\tilde{N}_T}{2|{q}^2-{\omega}^2
  +{\Pi}_T|^2}\nonumber\\
&+&{\rm Re}\frac{\tilde{N}_{LT}}{({q}^2+{\Pi}_L)({q}^2-{\omega}^2
+{\Pi}_T)^*}\Bigg\}.\qquad
\eea

\section{Numerical results}
\label{sec:results}

We performed numerical calculations of the relaxation times using
Eqs.~\eqref{eq:X-}, \eqref{eq:X+} and
\eqref{eq:nu_ei1}--\eqref{eq:nu_pe1}, and the electrical
conductivities from Eq.~\eqref{eq:sigma1} in the regime relevant to
neutron star outer crusts, considering a plasma composed of a single
ion species across the entire temperature–density range
studied. Below, we first present the phase structure of crustal
matter, followed by a discussion of our new results for relaxation
times and conductivities. A more detailed discussion including
  the phase structure of the crust and relaxation times is provided
  for $\isotope[12]{C}$ nuclei, but the final results for
  conductivities are also presented for $\isotope[56]{Fe}$ and
  $\isotope[4]{He}$ nuclei. For practical astrophysical applications,
all numerical values are given in in centimeter-gram-second units.

\subsection{Phase diagram of crustal plasma}

\begin{figure}[t] 
\begin{center}
\includegraphics[width=8cm,keepaspectratio]{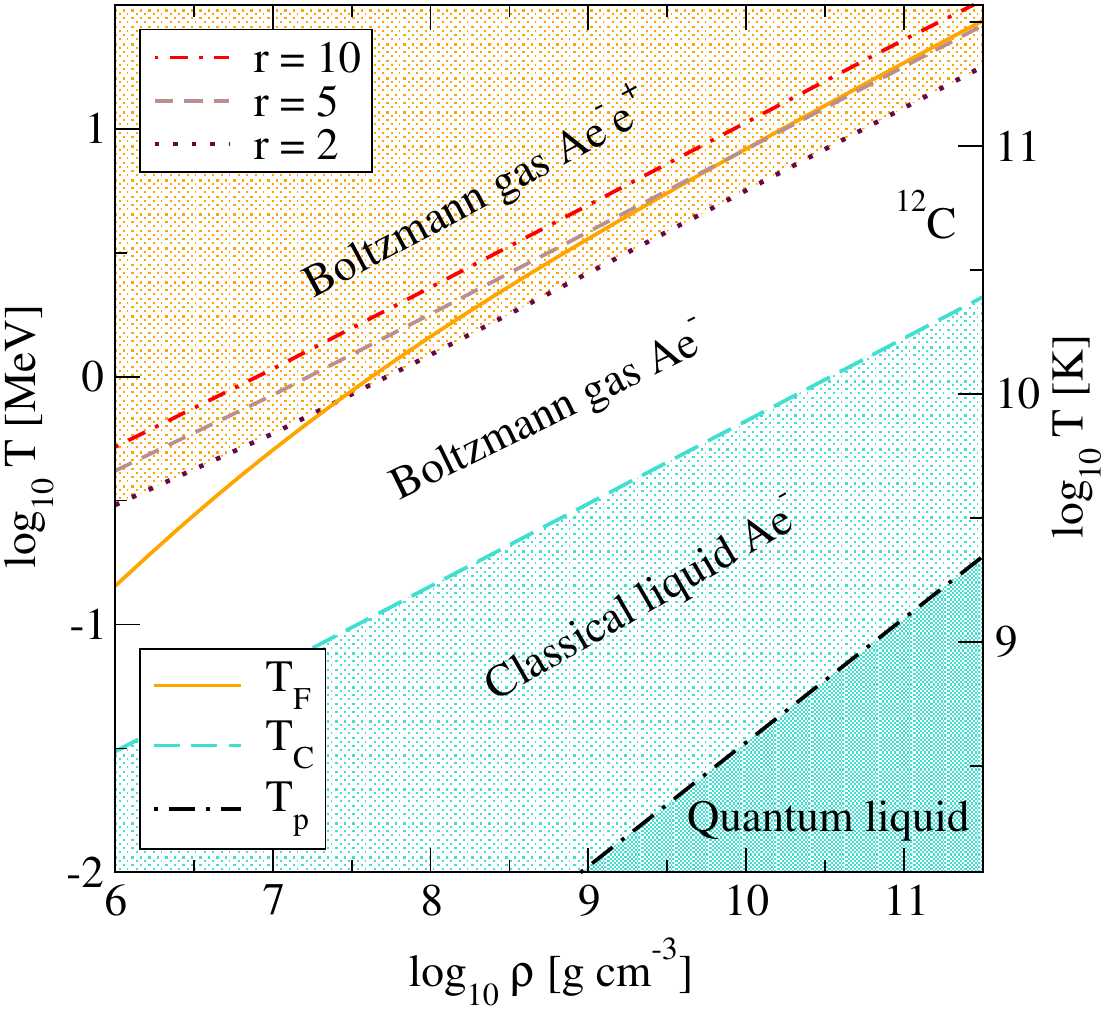}
\caption{ Temperature-density phase diagram of crustal plasma
  composed of carbon.  The ionic component of plasma forms a Boltzmann gas
  above the Coulomb temperature $T_C$, a classical liquid at
  $T_p\leq T\leq T_C$, and a quantum liquid below the plasma
  temperature $T_p$.  Electrons become degenerate below the Fermi
  temperature $T_F$. The three curves around $T_F$ correspond to
  temperatures where the ratio of the total electron density $n^-$ to
  the net electron density $n_e$ reaches the value $r$. }
\label{fig:PhaseDiagram} 
\end{center}
\end{figure}
\begin{figure}[t] 
\begin{center}
\includegraphics[width=7.8cm,keepaspectratio]{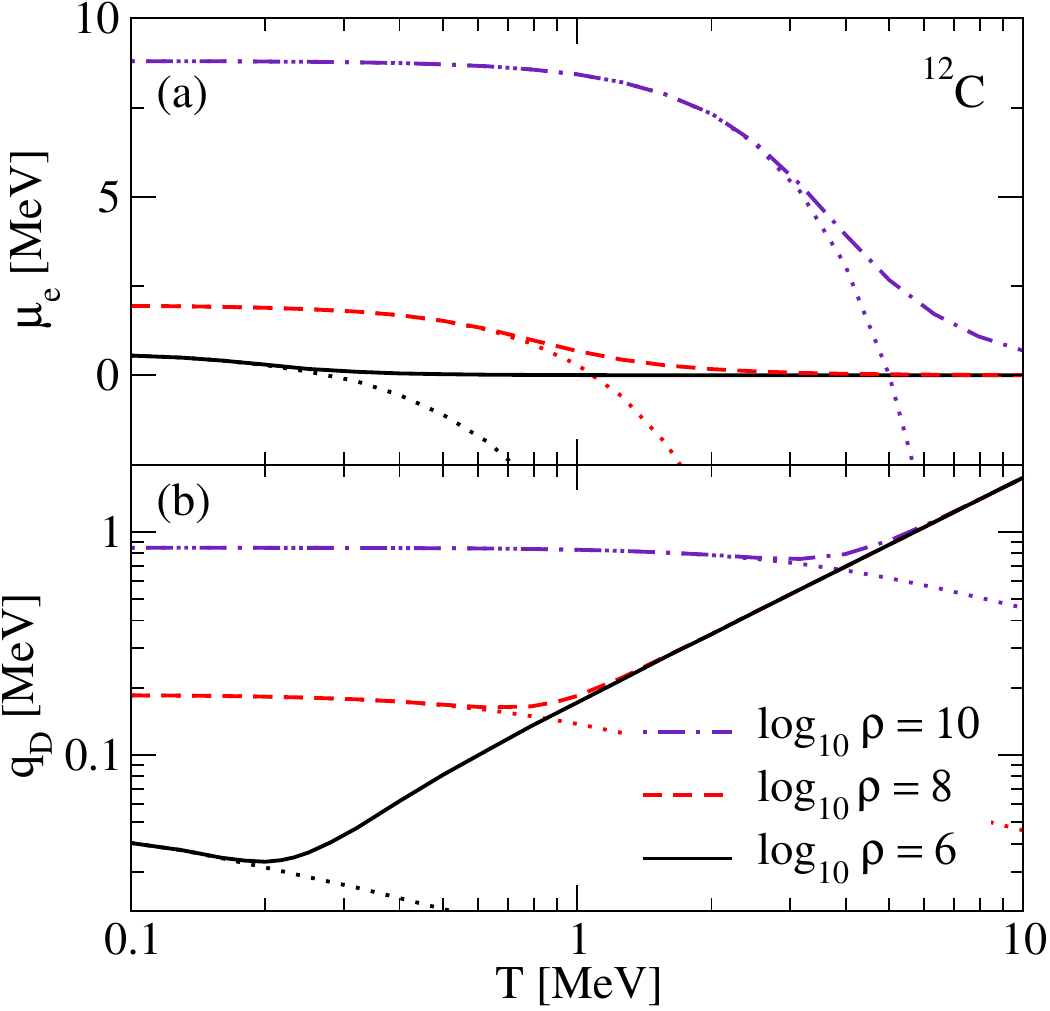}
\caption{Electron chemical potential $\mu_e$ (panel a) and the
  Debye wave number $q_D$ (panel b) as functions of the temperature
  for three values of the density indicated in the plot. The dotted
  lines show the same quantities in the case where positrons are neglected.}
\label{fig:mu_qd} 
\end{center}
\end{figure}

To describe the state of the crustal plasma, we introduce the
following physical quantities. The ion mass is defined as $M=Am_n$,
where $A$ is the ion mass number and $m_n\simeq 939$~MeV is the
average nucleon mass. The ion number density $n_i$ is related to the
density of matter $\rho$ by $n_i=\rho/M$, and we also define the
radius of the spherical volume per ion (the so-called Wigner Seitz
cell~\cite{Shapiro1983}) as $a=(4\pi n_i/3)^{-1/3}$.

The state of ions with charge number $Z$ is determined by the
value of the Coulomb plasma parameter
\bea\label{eq:Gamma}
\Gamma=\frac{(Ze)^2}{aT},
\eea
which is the ratio of the ion-ion Coulomb repulsion energy to their thermal energy. 
If $\Gamma\ll 1$ or,
equivalently $T\gg T_{\rm C}\equiv (Ze)^2/a$, ions are weakly coupled and  form a Boltzmann gas. In the regime where $\Gamma\ge 1$, ions are strongly coupled and form a liquid for values
$\Gamma\leq\Gamma_m\simeq 160$ and a solid for $\Gamma>\Gamma_m$. The melting temperature of the lattice associated
with $\Gamma_m$ is defined as $T_m=(Ze)^2/\Gamma_ma$. The plasma temperature
\bea 
T_p = \biggl(\frac{4\pi  Z^2e^2n_i}{M }\biggl)^{1/2}
\eea
determines the regime where the collective quantum effects in the plasma become important.  

Figure~\ref{fig:PhaseDiagram} shows the temperature–density phase diagram of the crustal plasma composed of carbon $\isotope[12]{C}$. The melting temperature lies below the plasma temperature and is therefore not shown in the diagram. This indicates that, as the temperature decreases, quantum effects in the carbon plasma become significant before the matter solidifies. The present study focuses on the portion of the phase diagram above the plasma temperature.

\begin{figure*}[t] 
\begin{center}
\includegraphics[width=7.75cm,keepaspectratio]{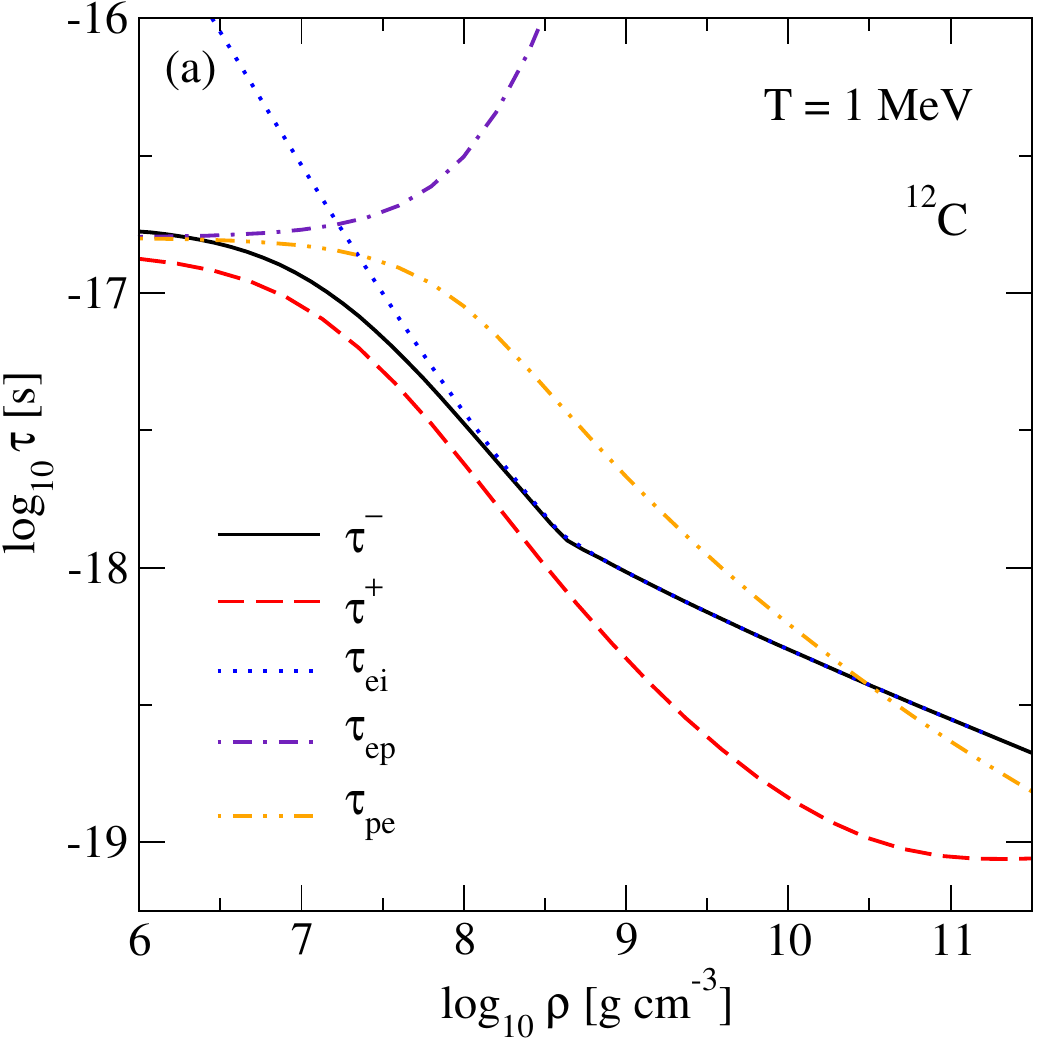}
\hspace{1cm}
\includegraphics[width=8cm,keepaspectratio]{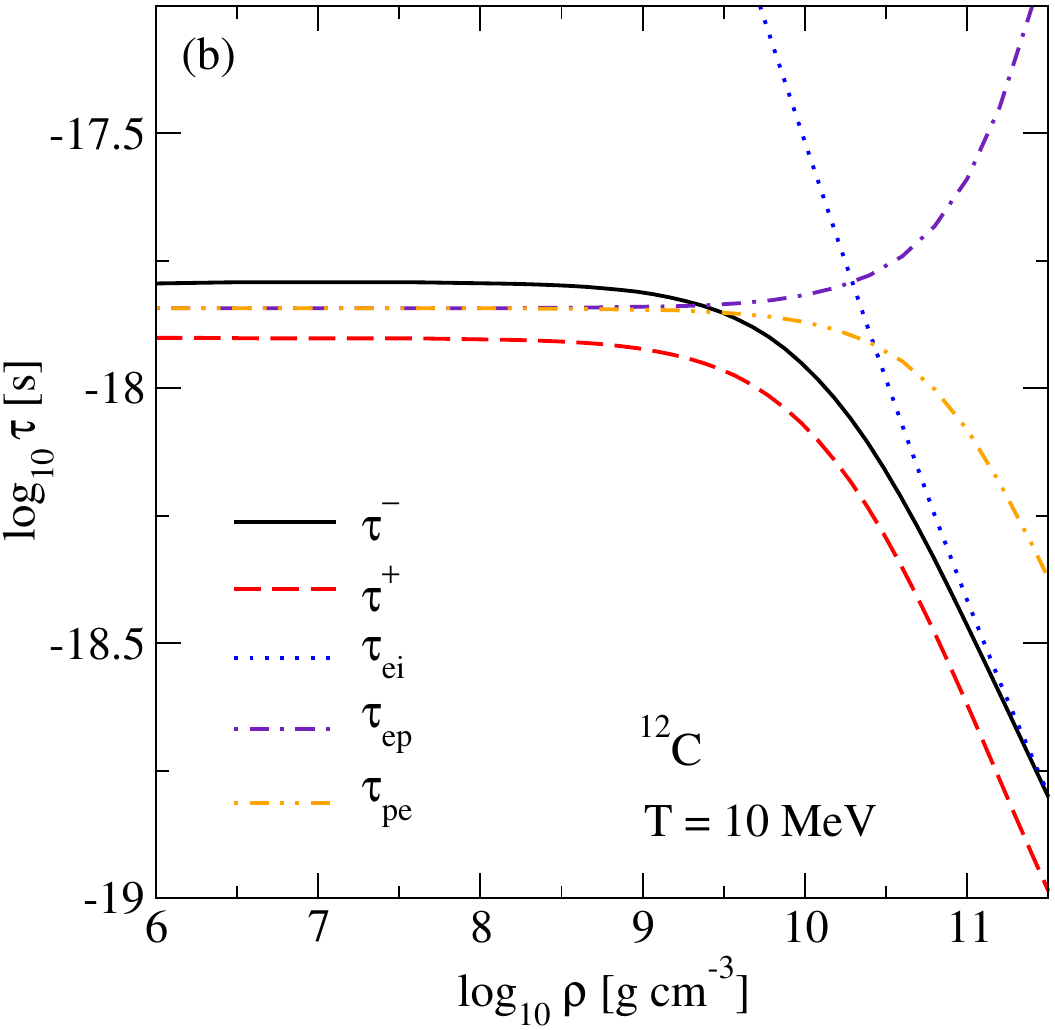}
\caption{ Dependence of electron (solid lines) and positron (dashed lines) relaxation times on density for the temperature fixed at (a) $T=1$~MeV; and (b) $T=10$~MeV. The dotted lines show the electron-ion relaxation time $\tau_{ei}=\ep/\nu_{ei}$, the dash-dotted lines show the electron-positron relaxation time $\tau_{ep}=\ep/2\nu_{ep}$, and the dash--double-dotted lines show the positron-electron relaxation time $\tau_{pe}=\ep/2\nu_{pe}$.}
\label{fig:tau_dens}
\end{center}
\end{figure*}

Electrons become degenerate below the Fermi temperature
$T_F = \ep_F -m$, where the Fermi energy is defined as
$\ep_F= (p_F^2+m^2)^{1/2}$, and the Fermi momentum is given by
$p_F = (3\pi^2n_e)^{1/3}$. At finite temperatures, 
once the thermal energy of matter becomes sufficient to allow for
the creation of electron-positron pairs, a thermal population of positrons
appears.  The electron number excess over positrons, $n_e$ is now
the relevant quantity that determines the charge neutrality condition
$n_e=Zn_i$.  The electron chemical potential is found
from the value implied by the charge conservation for electron excess
$n_e =n^{-} - n^{+}$ for any given values of $\rho$ and $T$.  The
positron and electron proper number densities are given by
\bea\label{eq:n_pm}
n^{\pm} = \int\!\frac{2d\bm p}{(2\pi)^3} f^{0\pm}(\ep)=\frac{1}{\pi^2}\! \int_m^{\infty}\! d\ep\,p\ep f^{0\pm}(\ep),
\eea
where $f^{0\pm}$ are the Fermi distributions of electrons $(-)$ and
positrons $(+)$ introduced earlier. Clearly,  $n^{-} > n_e$ when the population of positrons is not negligible.

In order to assess the relative importance of positrons in the plasma,
we define the ratio $r=n^{-}/n_e$. Physically, $r$ is expected to rise
rapidly with the temperature as a result of the fast opening of
kinematic phase space for the pair creation with increasing $T$. This
feature is clearly seen in Fig.~\ref{fig:PhaseDiagram}, where we have
plotted three lines which correspond to the values $r=2, 5, 10$.
These curves highlight the crucial importance of including
electron–positron pair creation processes in the composition of
crustal matter, even at temperatures near the Fermi temperature. The
ratio $r$ also increases with decreasing density at fixed $T$, as the
pair creation rate is closely related to the ratio $T/T_F$. The shaded
region at the top of the phase diagram corresponds to the regime where
the number of created positrons (or, equivalently, that of created
pairs), $n^+$, already exceeds the number of electrons in the cold
$(T \ll T_F)$ plasma. At temperatures $T \gg T_F$, the number of pairs
increases approximately cubically with temperature, as implied by
Eq.~\eqref{eq:n_pm}, yielding $n^+ \simeq n^- \gg n_e$ in this regime.

We also show the electron chemical potential $\mu_e$ and the Debye
wave number in Fig.~\ref{fig:mu_qd} as functions of temperature for
three values of the density. The dotted lines correspond to the case
in which the positron abundance is neglected. In the degenerate limit,
$T \ll T_F$, the positron contribution vanishes, and we recover
$\mu_e \simeq \epsilon_F$ and
$q_D^2 \simeq 4 e^2 p_F \epsilon_F / \pi$, which depend only on the
density, as seen in the figure. In the opposite, high-temperature
limit $T \gg T_F$, the chemical potential tends to zero,
$\mu_e \to 0$, and electrons and positrons contribute nearly equally
to the Debye wave number. In this regime, Eq.~\eqref{eq:Debye} yields
the approximation $q_D^2 \simeq 16 e^2 T^2/\pi.$ This behavior
contrasts with the high-temperature scaling
$q_D^2 \simeq 4 e^2 \pi n_e / T$ obtained when the positron
contribution is neglected. Consequently, the presence of positrons
leads to significantly larger, density-independent Debye wave numbers
and therefore to stronger screening of scattering amplitudes compared
to the positron-free case.

\subsection{Relaxation times}

\begin{figure*}[t] 
\begin{center}
\includegraphics[width=8cm,keepaspectratio]{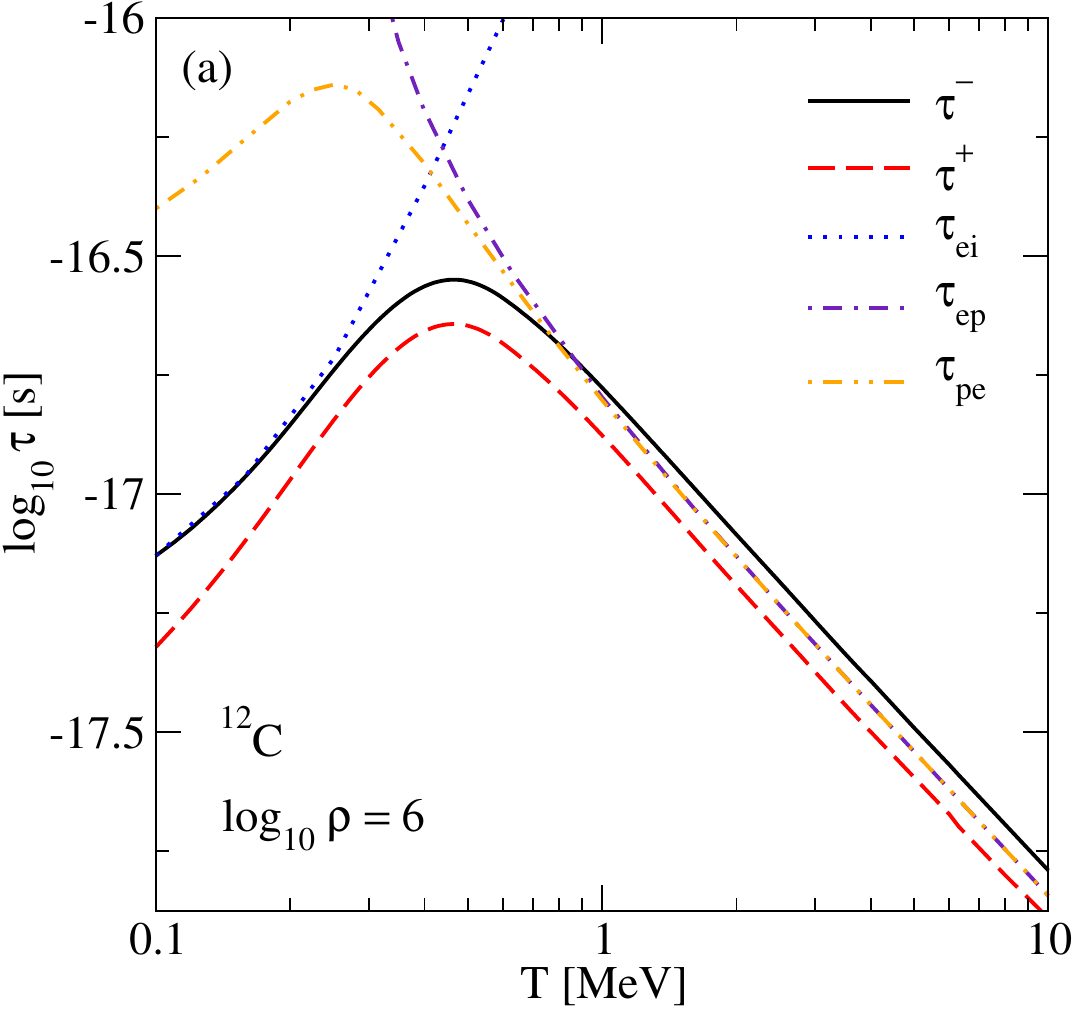}
\hspace{1cm}
\includegraphics[width=8cm,keepaspectratio]{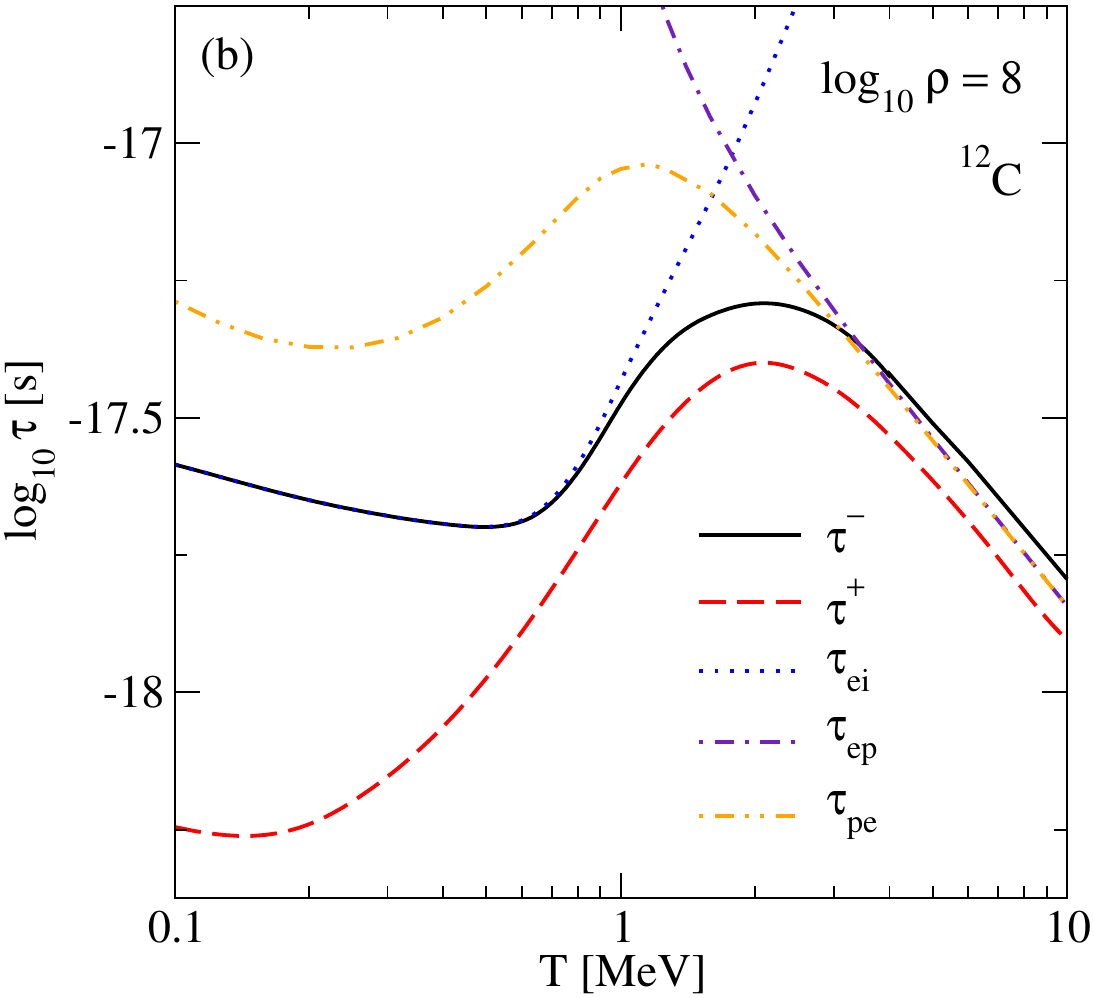}
\caption{Temperature dependence of electron (solid lines) and positron
  (dashed lines) relaxation times for the density fixed at (a)
  $\rho=10^6$~g\,cm$^{-3}$ and (b) $\rho=10^8$~g\,cm$^{-3}$. The
  dotted lines show the electron-ion relaxation time
  $\tau_{ei}=\ep/\nu_{ei}$, the dash-dotted lines show the
  electron-positron relaxation time $\tau_{ep}=\ep/2\nu_{ep}$, and the
  dash--double-dotted lines show the positron-electron relaxation time
  $\tau_{pe}=\ep/2\nu_{pe}$.  }
\label{fig:tau_temp}
\end{center}
\end{figure*}

 In this section, we discuss the electron and positron
  relaxation times arising from electron-ion, positron-ion, and
  electron-positron scattering processes.  Let us recall that the
  electron–positron and positron–electron collision rates given by
  Eqs.~\eqref{eq:nu_ep_final} and \eqref{eq:nu_pe_final} contain two
  distinct microscopic interaction channels: ordinary scattering and
  pair annihilation--creation processes. The latter is found to
  contribute significantly less to the total collision rate than the
  genuine scattering channel.

  Figure~\ref{fig:tau_dens} shows the density dependence of the
  relaxation times at fixed temperatures $T = 1$ MeV (left panel) and
  $T = 10$ MeV (right panel). Since the relaxation times are energy
  dependent $(\tau^{\pm} = \epsilon X^{\pm}$ with
  $X^{\pm} = \mathrm{const})$, their evaluation depends on the
  degeneracy of the electrons. To distinguish between the degenerate
  and nondegenerate regimes, we introduce the characteristic
  temperature $T^* \equiv T_F/3$, which serves as the transition scale
  between the two regimes. In degenerate matter, $T \le T^*$, the
  electron relaxation time $\tau^{-}$ is evaluated at the electron
  Fermi energy $\epsilon_F$, while in the nondegenerate
  ultrarelativistic regime, $T \ge T^*$, it is evaluated at the
  thermal energy $\bar{\epsilon} = 3T$.  The positron relaxation time
  $\tau^{+}$, by contrast, is always evaluated at the thermal energy.

  In order to assess the relative importance of the separate
  scattering processes, we show the partial contributions to the
  relaxation times due to electron-ion and electron-positron
  collisions. At high densities, where the positrons are irrelevant,
  the electron relaxation times coincide with those due to
  electron-ion collisions $\tau_{ei}\equiv \ep/\nu_{ei}$.  In this
  regime $\tau^{-}$ decreases with density because of increasing
  electron-ion scattering rates; here, we recover the old results of
  Ref.~\cite{Harutyunyan2016}.

In the opposite limit of dilute matter, where $T\gg T_F$,
electron-positron collisions strongly dominate $ei$ and $pi$
collisions, and $\tau^{-}$ and $\tau^{+}$ are close to their limiting
values $\tau_{ep}\simeq \tau_{pe}\equiv \ep/2\nu_{ep}$, as follows
from Eq.~\eqref{eq:tau_effective_ep}.  Here $\tau_{ep}$ ($\tau_{pe}$)
tends to a density-independent value below the transition point, but
increases (decreases) above that point, as the positron population
drops; the net electron density increases when matter enters the
degenerate regime.

The total electron relaxation time $\tau^{-}$ interpolates between
these two regimes, approaching a density-independent value $\tau_{ep}$
at small densities and decreasing at higher densities. The transition
point moves to higher densities for larger values of $T$. We remark
also that the longitudinal and transverse parts of the scattering
amplitudes contribute almost equally to the electron-positron
collision rates $\nu_{ep}$ and $\nu_{pe}$ in the regime where these
are relevant.

The behavior of the positron relaxation time $\tau^{+}$ is very
similar to that of $\tau^{-}$ but with slightly lower values. The
partial contribution $\tau_{pi}$ is very close to $\tau_{ei}$ in the
regime where it is relevant; therefore, it is not shown in the
figures.  Indeed, as indicated by Eqs.~\eqref{eq:l_pm} and
\eqref{eq:nu_sigma_ei_final}, the electron-ion and positron-ion
collision rates are essentially identical in the regime where
positrons are sufficiently abundant, since the integrals over
combinations of Fermi functions in the numerators and denominators
largely cancel.

Figure~\ref{fig:tau_temp} shows the temperature dependence of the
relaxation times for several fixed values of the density. We see again
that, in the strongly degenerate regime $T\ll T_F$, we have
$\tau^{-}\simeq \tau_{ei}$, which decreases with the temperature due
to the ion structure factor $S(q)$~\cite{Harutyunyan2016}.  At
intermediate temperatures $T \sim T^*$, where the matter is
semidegenerate, $\tau_{ei}$ changes its behavior from a decreasing to
an increasing function. This transition reflects the approximate
scaling $\tau_{ei} \propto \epsilon \propto T$. In the
semidegenerate regime, there exists a temperature window in which
electron–positron collisions remain subdominant, i.e.,
$\tau_{ei} \gtrsim \tau_{ep}$. As a result, the effective electron
relaxation time $\tau^{-}$ increases in this domain.  However, at
higher temperatures $T\gg T_F$, the electron-positron collisions
become dominant, and we have $\tau^{-}\simeq \tau_{ep}\propto T^{-1}$,
which follows from dimensional arguments. Thus, as a result of
nontrivial convolution of electron-ion and electron-positron
collisions, the electron relaxation time $\tau^{-}$ shows a
nonmonotonic behavior with temperature, being a decreasing function
in both limiting cases (with a higher slope in the nondegenerate
regime), and increasing in the transitional regime which interpolates
between these two limiting cases. The behavior of the positron
relaxation time $\tau^{+}$ is very similar to that of $\tau^{-}$ in
the transitional and nondegenerate regimes where it is relevant, 
albeit with slightly smaller values.  

\subsection{Electrical conductivity from electron-ion and positron-ion collisions}

\begin{figure}[t] 
\begin{center}
\includegraphics[width=8cm,keepaspectratio]{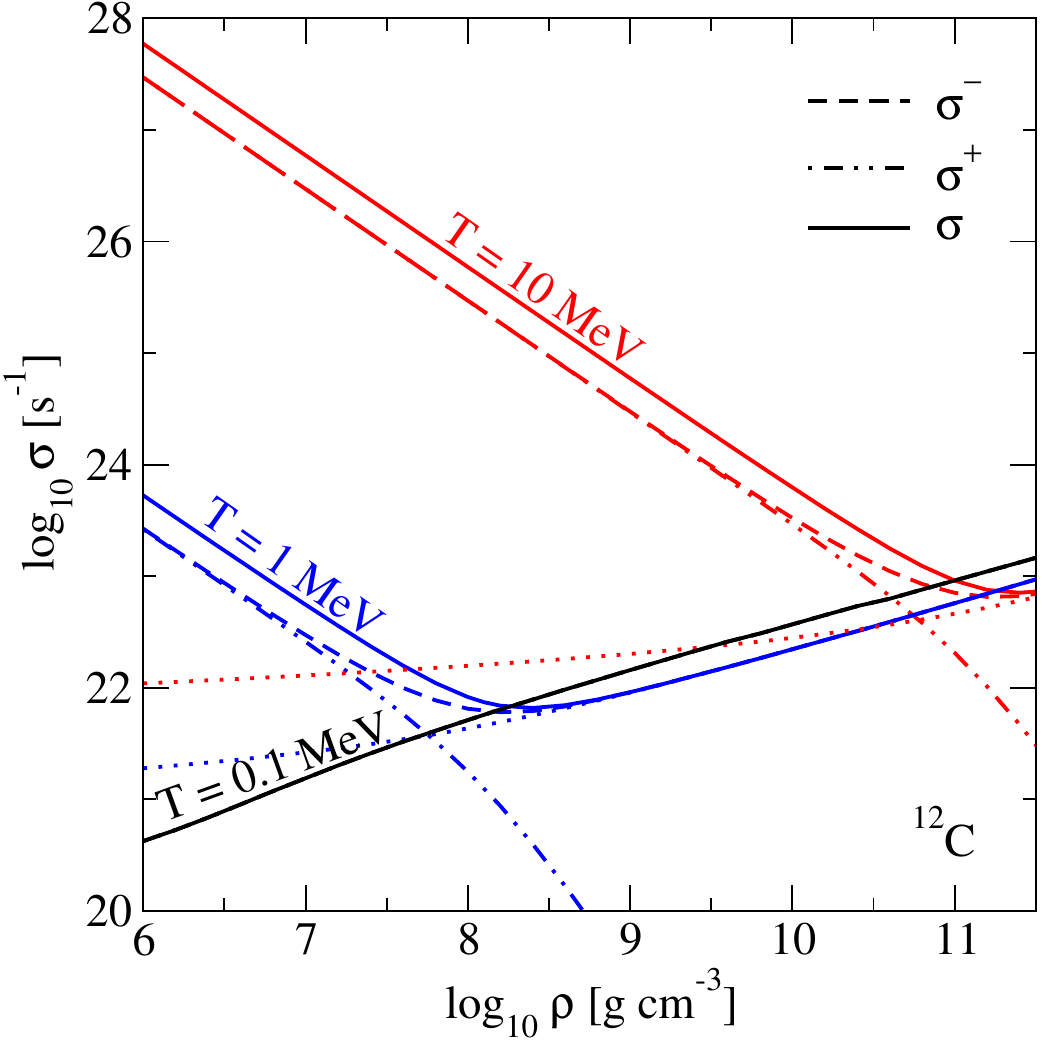}
\caption{ Dependence of electronic (dashed lines) and positronic (dash-double-dotted lines) partial conductivities and their sum (solid lines) for three values of the temperature indicated in the plot. These are the conductivities resulting only from $ei$ and $pi$ collisions. The dotted lines show the conductivity from Ref.~\cite{Harutyunyan2016} where positron contribution is neglected.}
\label{fig:sigma_dens}
\end{center}
\end{figure}
\begin{figure}[t] 
\begin{center}
\includegraphics[width=8cm,keepaspectratio]{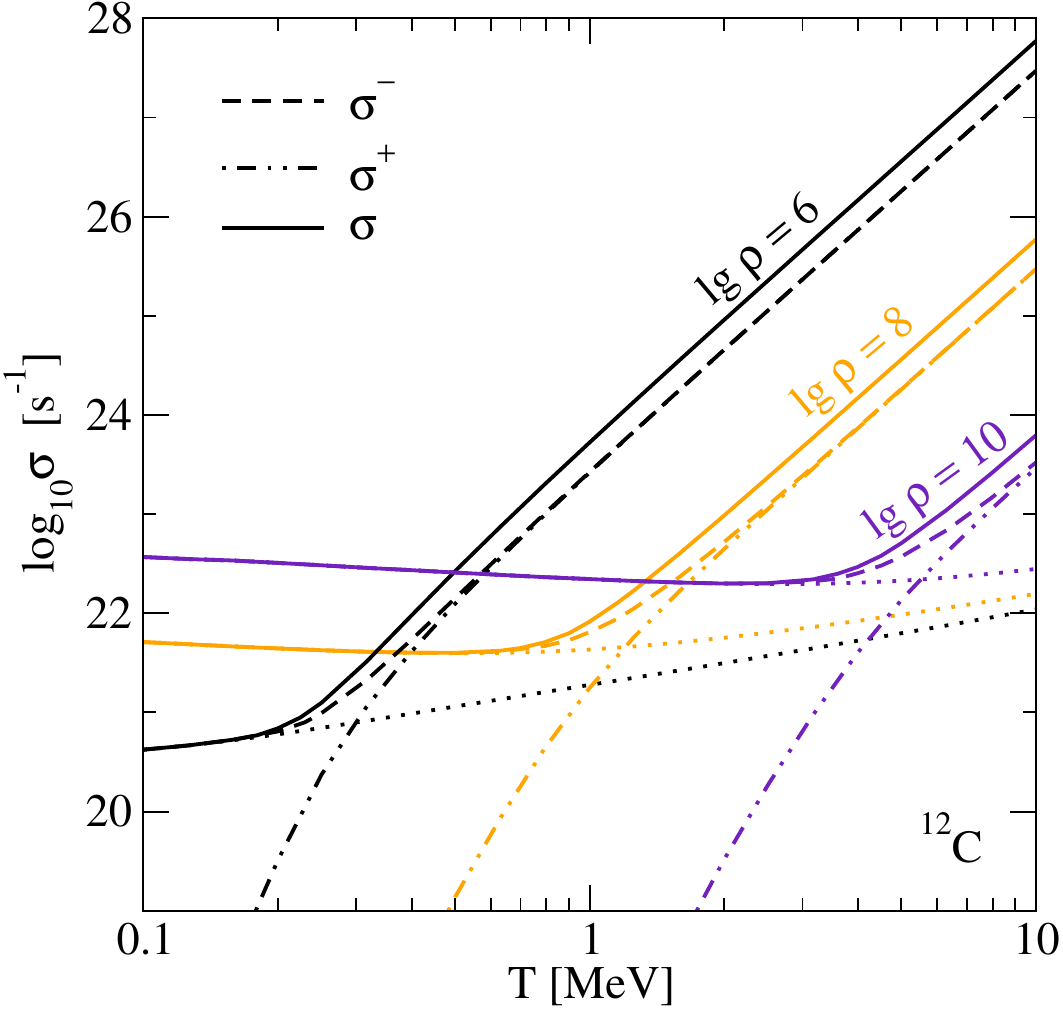}
\caption{ Temperature dependence of electronic (dashed lines) and positronic (dash-double-dotted lines) partial conductivities and their sum (solid lines) for three values of the density indicated in the plot. These are the conductivities resulting only from $ei$ and $pi$ collisions. The dotted lines show the conductivity from Ref.~\cite{Harutyunyan2016} where positron contribution is neglected.
}
\label{fig:sigma_temp}
\end{center}
\end{figure}

For clarity, we first consider the conductivities that include only
electron–ion and positron–ion collisions. The full conductivities,
which also account for electron–positron collisions, will be addressed
in the next subsection. As discussed above,
$\tau_{ei}\simeq \tau_{pi}$ in the regime where positrons are
sufficiently abundant; therefore, the primary distinction between the
electron and positron contributions to the electrical conductivity
arises from the Fermi distribution functions in Eq.~\eqref{eq:sigma1}.

Let us first analyze the formula~\eqref{eq:sigma1} in the limiting
cases of strongly degenerate and nondegenerate electrons.  In the
degenerate limit where $T\ll T_F$, the positron contribution to the
conductivity is suppressed, and $\sigma^-$ can be simplified via the
substitution $\partial f^{0-}/\partial\ep\to -\delta(\ep-\ep_F)$,
which leads to $l^{\pm}\to p_F^3 T/2\pi^2$ and, therefore (recall that
$X^{\pm}=\tau^{\pm}/\ep$)
\bea\label{eq:sigma_deg}
\sigma\simeq\frac{n_ee^{2}\tau_F}{\ep_F},
\eea
where $\tau_F\equiv\tau^-(\ep_F)$. This is the well-known Drude
formula for the electrical conductivity in the theory of
metals~\cite{Abrikosov:Fundamentals}.

In the opposite limit of high temperatures $T\gg T_F$, we also have $T\gg\mu_e$, 
electrons and positrons are ultrarelativistic with the average thermal
energy $\bar{\ep}\simeq 3T$, and their number densities
are almost equal, $n^{+}\simeq n^{-}\equiv \bar{n}\sim \bar{\ep}^3$.
In this case, $f^{0+}\simeq f^{0-}\equiv f^{0}\ll 1$, and
Eq.~\eqref{eq:sigma1} can be approximated as
\bea\label{eq:sigma_non_deg}
\sigma^{+} \simeq \sigma^{-}
\simeq \frac{\bar{n}e^{2}\bar{\tau}}{\bar{\ep}},
\eea
where $\bar{\tau}=\tau(\bar{\ep})$, and we used the definition of the
statistical average of an energy-dependent quantity $F(\ep)$ as
\bea\label{eq:average}
\bar{n} \bar{F} 
=\frac{1}{\pi^2}\int_0^\infty {p^2dp}\, F(\ep)f^{0}(\ep).
\eea
Thus, the formulas for conductivity in both strongly degenerate and
nondegenerate regimes have the same form but contain different energy
scales, specifically, $\ep_F$ in the degenerate regime and
$\bar{\ep}\simeq 3T$ in the nondegenerate, ultrarelativistic regime.

The relaxation time $\eqref{eq:nu_sigma_ei_final}$ in the physical
regimes of interest was discussed in detail in
Ref.~\cite{Harutyunyan2016}. On average, it was found that, for
carbon, $\tau$ varies with density and temperature according to the
scaling $\tau(\epsilon) \propto
\epsilon^{2}\rho^{-0.9}T^{-0.2}$. Using this scaling and the rough
estimates $n_e \sim \epsilon_F^{3}$ and
$\bar{n} \sim \bar{\epsilon}^{3}$, Eqs.~$\eqref{eq:sigma_deg}$ and
$\eqref{eq:sigma_non_deg}$ yield a universal scaling
$\sigma \propto \epsilon^{4}\rho^{-0.9}T^{-0.2}$. This corresponds
approximately to $\sigma \propto \rho^{0.5}T^{-0.2}$ in the degenerate
regime and $\sigma \propto \rho^{-1}T^{4}$ in the nondegenerate
regime.

Figure~\ref{fig:sigma_dens} shows the partial conductivities
$\sigma^-$ and $\sigma^+$, as well as the total conductivity
$\sigma=\sigma^-+\sigma^+$ as functions of density for three
temperature values, which are chosen to cover the range from the
degenerate regime ($T=0.1$~MeV) to the nondegenerate regime
($T=10$~MeV). The intermediate value $T=1$~MeV is a representative of
the transition between these regimes, which occurs around
$\rho \simeq 10^8$~g\,cm$^{-3}$; see Fig.~\ref{fig:PhaseDiagram}.  For
comparison, we also plot the conductivities obtained previously in
Ref.~\cite{Harutyunyan2016}, where the presence of positrons in 
matter was not taken into account.

\begin{figure}[t] 
\begin{center}
\includegraphics[width=8cm,keepaspectratio]{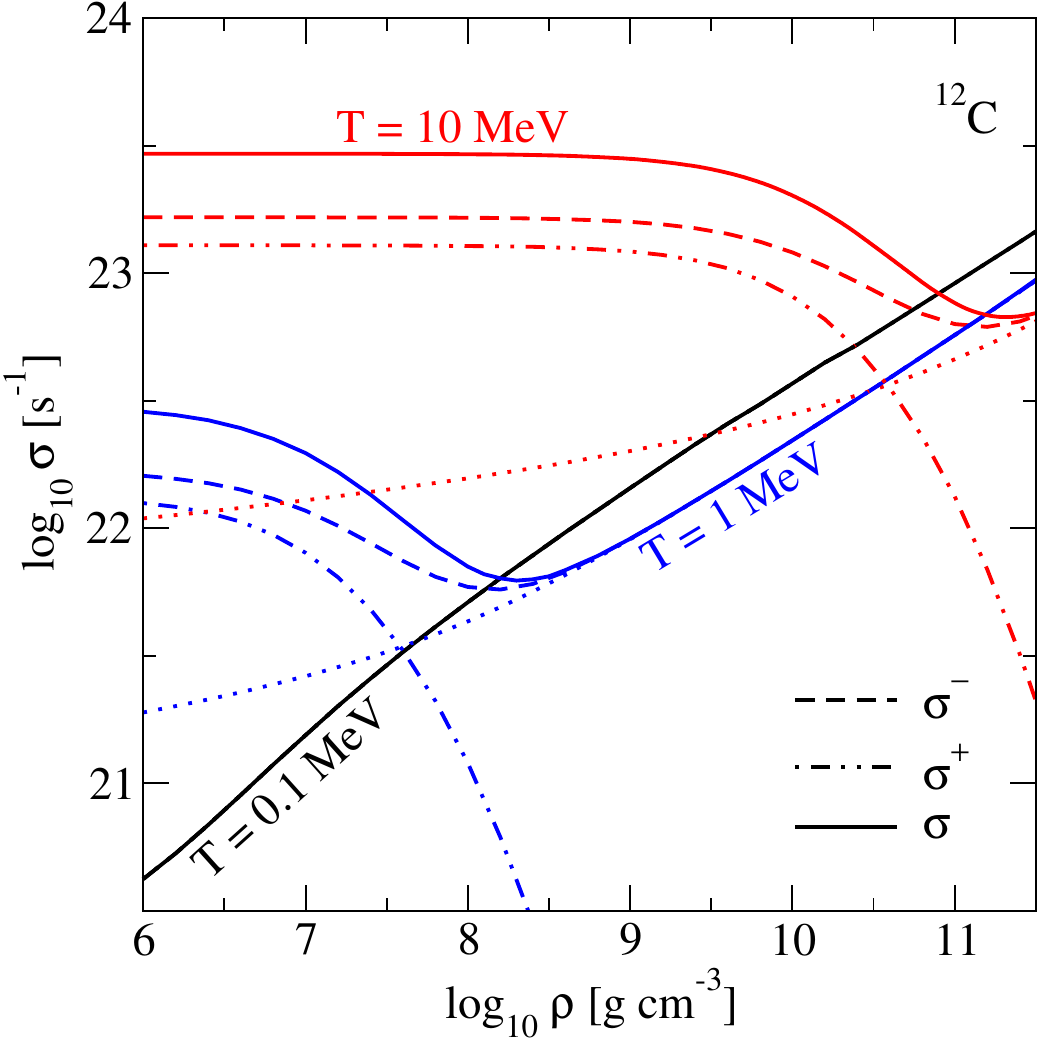}
\caption{ Same as Fig.~\ref{fig:sigma_dens}, but with the inclusion of $ep$ collisions. }
\label{fig:sigma_dens_ep}
\end{center}
\end{figure}
\begin{figure}[t] 
\begin{center}
\includegraphics[width=8cm,keepaspectratio]{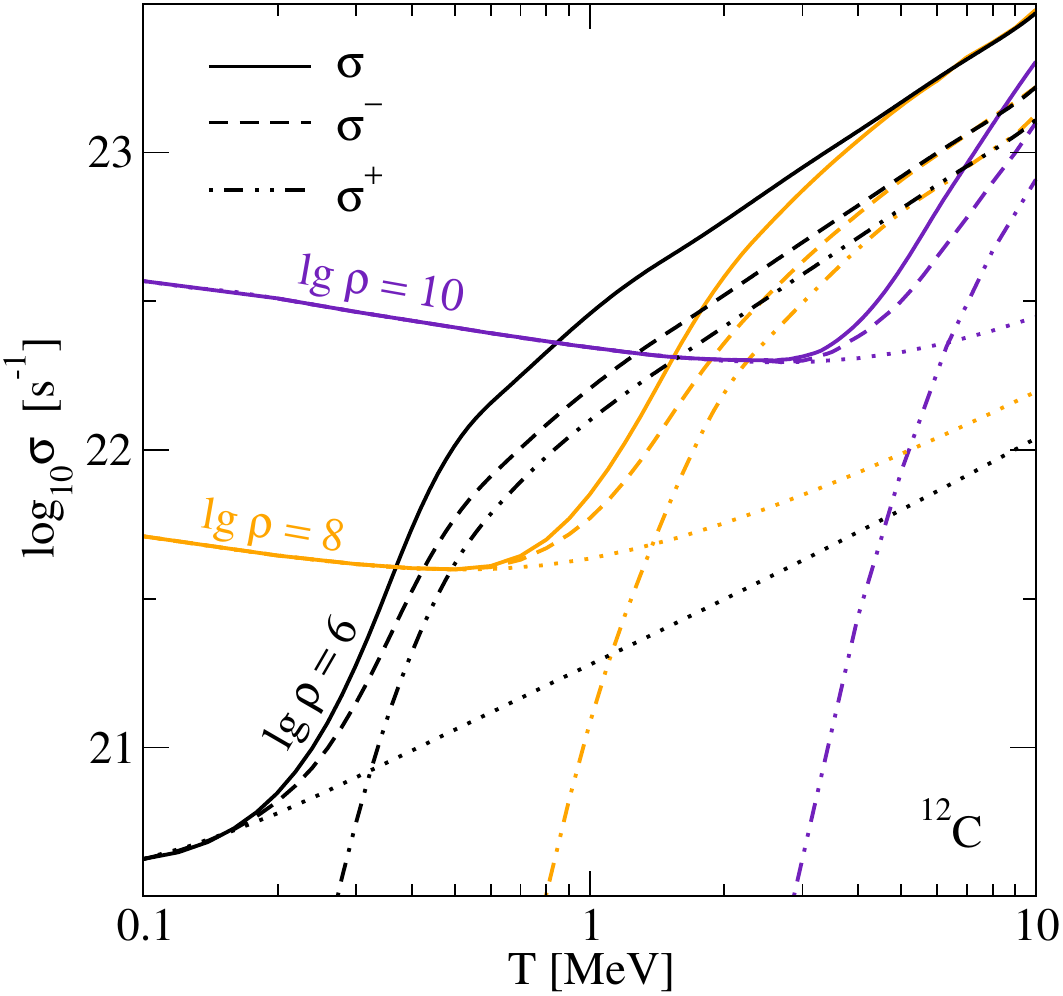}
\caption{ Same as Fig.~\ref{fig:sigma_temp}, but with the inclusion of $ep$ collisions. 
}
\label{fig:sigma_temp_ep}
\end{center}
\end{figure}

The qualitative behavior of the conductivity in the degenerate regime
is opposite to that in the transitional and nondegenerate
regimes. This contrast can be understood by noting that at a
temperature of $T = 0.1$ MeV, positrons are effectively absent since
the temperature is far below the electron rest mass. The positron
distribution is also strongly suppressed at
$\rho \ge 10^{8}\,\mathrm{g\,cm^{-3}}$ for $T = 1$ MeV due to Pauli
blocking: As the Fermi energy increases, the number of kinematically
accessible states for pair creation diminishes. In both situations,
the present results for $\sigma$ coincide with the earlier
work~\cite{Harutyunyan2016}, as expected. In this regime, the
conductivity scales as $\sigma \propto \rho^{0.5}$, because the
increase in the electron density of states with $\rho$ outweighs the
decrease in $\tau_F$.

Positrons are seen to make a significant contribution to the
conductivity in the low-density regime
$\rho \le 10^{8}\,\mathrm{g\,cm^{-3}}$ for $T = 1$ MeV, and across
nearly the entire outer crust
($\rho \le 10^{11}\,\mathrm{g\,cm^{-3}}$) for $T = 10$ MeV. In this
regime, the behavior of the conductivity reverses: At fixed
temperature, it decreases with density, following the power law
$\sigma \propto \tau \propto \rho^{-1}$. This occurs because in this regime, the total
number of electron–positron pairs, $\bar{n}$, is determined by
temperature alone and is independent of the matter density.

The two conductivity branches representing the degenerate and
nondegenerate regimes merge smoothly at a minimum located at their
intersection. This transition point corresponds approximately to the
temperature $T \simeq T^*$, where the two characteristic energy scales
become comparable.

Let us now turn to the temperature dependence of the conductivity (see
Fig.~\ref{fig:sigma_temp}). The low-temperature branches again
reproduce the results of Ref.~\cite{Harutyunyan2016}, decreasing
slowly with temperature due to the similar behavior of $\tau_F$, which
is governed by ion–ion correlations. In Ref.~\cite{Harutyunyan2016},
the conductivity was found to reach a minimum at the transition point
$T \simeq T^*$, resulting from the reversed temperature dependence of
the relaxation time in the nondegenerate regime.

As seen from Fig.~\ref{fig:sigma_temp}, the position of the minimum
around $T\simeq T^*$ also remains unchanged when positrons are
present. However, the slope of the curves above the transition point
rises abruptly, which is the consequence of the fast opening of the
phase space for the pair creation with increasing temperature, as seen
also from the phase diagram in Fig.~\ref{fig:PhaseDiagram}.  In
contrast to the previous scaling $\sigma\propto T^{0.8}$ in the
nondegenerate regime, here we find $\sigma\propto T^4$.  To give a
quantitative example, we compare the results with and without
positrons at the lowest density considered here,
$\rho = 10^{6}\,\mathrm{g\,cm^{-3}}$. The inclusion of positrons
increases the conductivity roughly by 2 orders of magnitude at $T = 1$
MeV and by 6 orders of magnitude at $T = 10$ MeV.

\subsection{Electrical conductivity including electron-positron collisions}

We now discuss the electrical conductivities obtained in the full
calculation, where electron–positron collisions are included as well.
Figure~\ref{fig:sigma_dens_ep} shows the partial conductivities
$\sigma^-$ and $\sigma^+$ and their sum $\sigma$ as functions of
density for the same temperature values as in
Fig.~\ref{fig:sigma_dens}.  Comparing these two figures, we see that
the effect of electron-positron collisions becomes crucial in the
low-density and high-temperature regime of nondegenerate matter where
$T\gg T_F$, as expected. Here, the conductivity becomes density
independent, as a consequence of density independence of relaxation
times in this regime and the relation~\eqref{eq:sigma_non_deg}. In the
degenerate regime, the electron-positron collisions are suppressed,
and we recover the results of the previous section. Although the
location of the minimum in the total conductivity remains unaffected
by electron-positron collisions, these are non-negligible already at
$T\geq T^*$.

\begin{figure}[t] 
\begin{center}
\includegraphics[width=8cm,keepaspectratio]{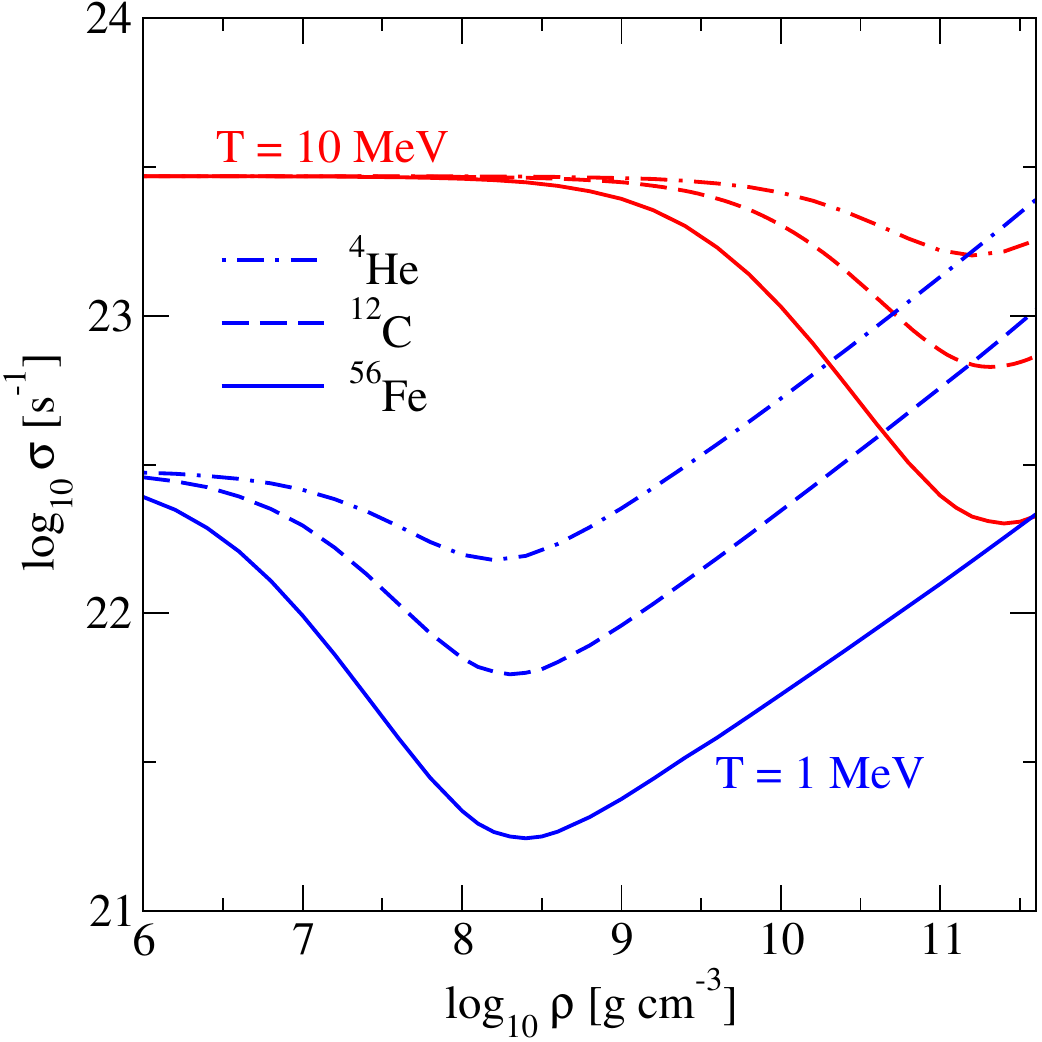}
\caption{ Total conductivity as a function of density for three different ionic compositions and two values of the temperature indicated in the plot.}
\label{fig:sigma_dens_comp}
\end{center}
\end{figure}
\begin{figure}[t] 
\begin{center}
\includegraphics[width=8cm,keepaspectratio]{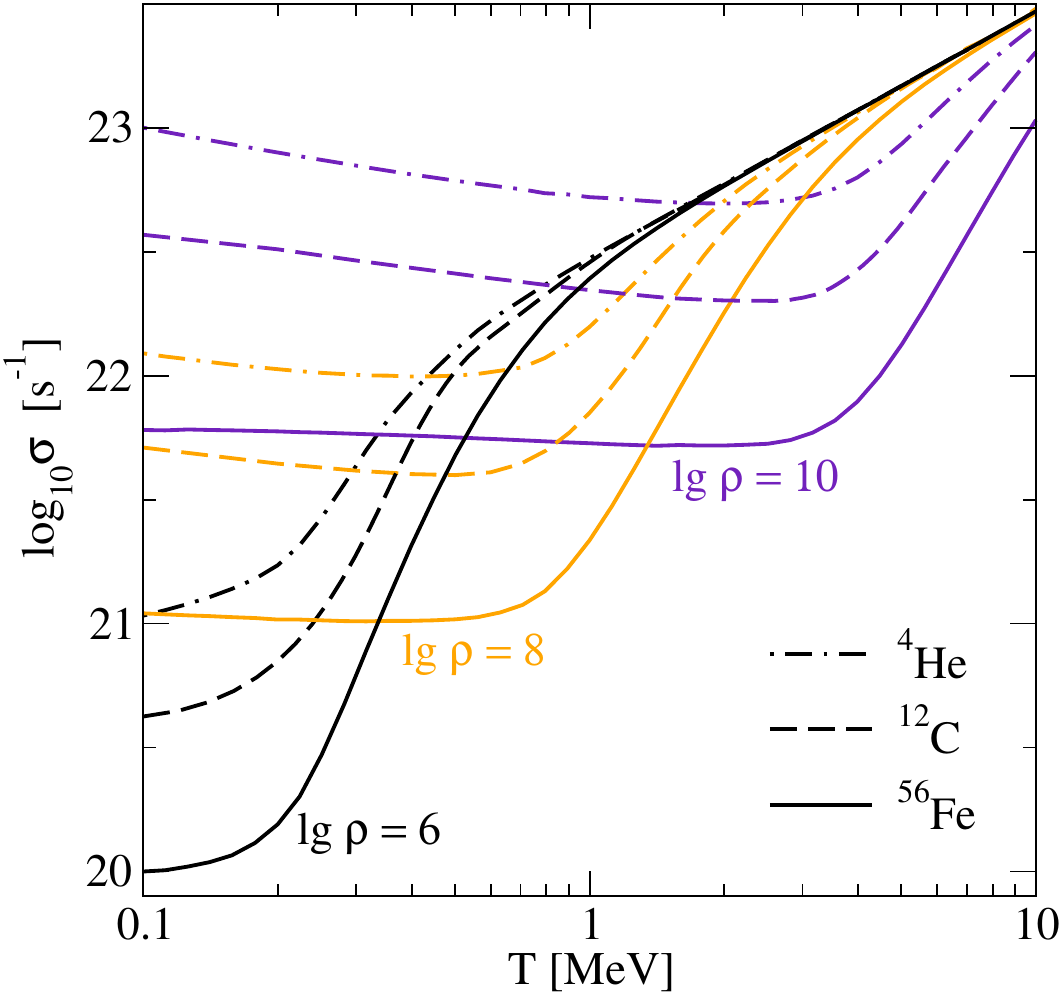}
\caption{ Total conductivity as a function of temperature for three different ionic compositions and three values of the density indicated in the plot.}
\label{fig:sigma_temp_comp}
\end{center}
\end{figure}

Turning to the temperature dependence of conductivities shown in
Fig.~\ref{fig:sigma_temp_ep}, we see again that the main difference as
compared to Fig.~\ref{fig:sigma_temp} appears in the high-temperature
regime above the minimum point $T\simeq T^*$.  At very high
temperatures, where $T \gg T_F$, the conductivity $\sigma$ becomes
essentially independent of density. On dimensional grounds, one then
finds the scaling $\sigma \propto T$ in this regime, which agrees with
the high-$T$ asymptotics shown in Fig.~\ref{fig:sigma_temp_ep}. This
behavior is particularly evident at $\rho = 10^6~\mathrm{g\,cm^{-3}}$.
Immediately above the transition temperature $T\simeq T^*$, $\sigma$
follows the power-law dependence $\sigma\propto T^4$, as the
collisions with ions are still dominant here. However, this fast
increase is mitigated at higher temperatures as a result of intense
electron-positron collisions, which surpass the $ei$ and $pi$
collision rates. As a numerical example, at $\rho=10^6$~g\,cm$^{-3}$,
the ratios of conductivities we find to those computed earlier in
Ref.~\cite{Harutyunyan2016} are factors of $20$ and $30$ for $T=1$ and
$T=10$~MeV, respectively. These factors decrease at higher densities
but still reach nearly an order of magnitude across most of the outer
crust at $T = 10$ MeV.

To give an estimate of how our results would change depending on
  the crust composition, we show the density and temperature
  dependence of the total conductivity $\sigma$ for a crust composed
  of $\isotope[4]{He}$, $\isotope[12]{C}$ and $\isotope[26]{Fe}$ in
  Figs.~\ref{fig:sigma_dens_comp} and \ref{fig:sigma_temp_comp},
  respectively. The qualitative behavior of the conductivity is
  similar for these three elements. In all cases, there is a minimum
  (both in the density and temperature dependence) at the point where
  $T\simeq T^*$. At densities above or temperatures below this point,
  the conductivity is dominated by electron-ion collisions and scales
  as $\sigma\propto Z^{-1}$, as follows from
  Eqs.~\eqref{eq:nu_sigma_ei_final} and \eqref{eq:sigma_deg} (recall
  that $n_e=Zn_i=Z\rho/Am_n$ is essentially independent of $Z$). In
  the high-temperature and low-density regime, the conductivity
  becomes independent of both density and nuclei species, approaching
  a temperature-dependent value
  $\sigma\,[{\rm s}]\simeq 3\times 10^{22}\, T\,[{\rm MeV}]$. Thus, the
  composition dependence of conductivity is less pronounced in the
  high-temperature regime where electron-positron collisions play the
  main role in relaxation processes.  

Concluding this section, we emphasize that accounting for
creation of positrons is essential for the 
accurate determination of the
transport properties in the outer crust 
of neutron stars, already at
temperatures close to the Fermi temperature.

\section{Summary}
\label{sec:summary}

In this work, we computed the electrical conductivity of hot
QED plasma relevant to the outer crusts of neutron stars. The
temperatures and densities studied here cover the transition from the
degenerate to the nondegenerate regime for electrons. The novelty of
our work is the inclusion of the contribution of positrons to the
conductivity, which becomes
increasingly  important as the temperature increases. In this regime, the conductivity is dominated by thermally created electron-positron pairs, which scatter off the correlated nuclei as well as off each other via screened electromagnetic force. 

Solving a system of coupled Boltzmann kinetic equations for electrons
and positrons, we expressed the conductivity in terms of electron and
positron relaxation times. These contain electron-ion, positron-ion
and electron-positron collision rates, which were computed by using
the relevant leading-order scattering diagrams, which include all
screening factors of strongly correlated crustal plasma.

Numerical results were obtained for conductivities mainly for
crustal matter composed of $\isotope[12]{C}$.  We found that the
conductivity increases with density in the degenerate regime, in
agreement with earlier results of Ref.~\cite{Harutyunyan2016}, but in
the nondegenerate regime, it shows a decreasing behavior, in contrast
to the results of previous studies. As a result, the conductivity
crosses a minimum at the transition point from the nondegenerate to
the degenerate regime. As a function of temperature, the conductivity
grows with a power law $\sigma\propto T^4$ in the semidegenerate
regime close to the transition temperature $T\simeq T_F/3$ because of
intense creation of thermal electron-positron pairs. In this region,
the number of pairs increases as $\propto T^3$, but the scattering is
still mainly due to ions. At higher temperatures $T\gg T_F$, the
electron-positron collisions become more important than electron-ion
and positron-ion collisions, resulting in a slower, almost linear
increase of $\sigma$ with temperature.  We find that the inclusion of
positrons enhances the electrical conductivity in the hot regions of
the outer crust by factors of about 10. The results obtained in this
work therefore underscore the importance of accounting for positrons
in the transport properties of heated plasma in neutron-star crusts.

The present study can be extended by considering more realistic,
temperature-dependent crustal compositions that include multiple
nuclear species in statistical equilibrium, as well as light nuclear
clusters. Such an approach would provide a more accurate description
of the microphysics in neutron star crusts, where the distribution of
nuclei can vary significantly with density and temperature. In
addition, the formalism developed here may be applied to improve
calculations of various transport coefficients, including thermal
conductivity, viscosities, and thermoelectric coefficients, in hot
neutron star matter. Incorporating the effects of positrons in these
calculations is particularly important, as they can significantly
modify screening and collision rates, thereby influencing the
transport properties of the warm crust. These extensions would allow
for a more comprehensive and realistic modeling of the thermal and
transport behavior of the hot transient state of compact stars at
subnuclear densities.

\section*{ACKNOWLEDGMENTS}

The authors acknowledge support from the Collaborative Research Grant
No. 24RL-1C010 provided by the Higher Education and Science Committee
(HESC) of the Republic of Armenia through the “Remote Laboratory”
program. A.~S. also acknowledges support from the Deutsche
Forschungsgemeinschaft Grant No. SE 1836/6-1 and the Polish National
Science Centre (NCN) Grant No. 2023/51/B/ST9/02798.

\begin{widetext}

\appendix

\section{Electron-positron scattering amplitude}
\label{app:matrix}

In this appendix, we will provide the details of the calculation of
the spin-averaged electron-positron scattering amplitude.  From
Eqs.~\eqref{eq:amplitude_ep}--\eqref{eq:amplitude_ep_s}, we have
\bea\label{eq:matrix_squared}
\big|{\cal M}_{12\to 34}^{ep}\big|^2 &=&
\big|{\cal M}_{12\to 34}^{ep,t}\big|^2
+\big|{\cal M}_{12\to 34}^{ep,s}\big|^2 
-2{\rm Re}\left({\cal M}_{12\to 34}^{ep,t}
{\cal M}_{12\to 34}^{ep,s*}\right), \\
\label{eq:matrix_t_squared}
|{\cal M}^{ep,t}_{12\to 34}|^2 &=& |{\cal M}^t_L|^2+
|{\cal M}^t_T|^2-2{\rm Re}({\cal M}^t_L{\cal M}^{t*}_T),\\
\label{eq:matrix_s_squared}
|{\cal M}^{ep,s}_{12\to 34}|^2 &=& |{\cal M}^s_L|^2+
|{\cal M}^s_T|^2-2{\rm Re}({\cal M}^s_L{\cal M}^{s*}_T),\\
\label{eq:matrix_ts}
{\cal M}_{12\to 34}^{ep,t}
{\cal M}_{12\to 34}^{ep,s*} 
&=& {\cal M}^t_L {\cal M}^{s*}_L
+{\cal M}^t_T {\cal M}^{s*}_T
-({\cal M}^t_L {\cal M}^{s*}_T
+{\cal M}^t_T {\cal M}^{s*}_L),
\eea
with the $t$-channel (ordinary scattering) terms
\bea\label{eq:matrix_ep_t_chanel}
|{\cal M}^t_L|^{2} = 
\frac{J_0J_0^*J'_0J_0'^{*}}{|t_0|^2},\quad
|{\cal M}^t_T|^{2} = 
\frac{J_{i\perp}J_{k\perp}^*J_{i\perp}'J_{k\perp}'^*}
{|t_\perp|^2},\quad
{\cal M}^t_L{\cal M}_T^{t*} =
\frac{J_0J_{i\perp}^*J_0'J_{i\perp}'^*}{t_0 t_\perp^*},
\eea
the $s$-channel (annihilation) terms
\bea\label{eq:matrix_ep_s_chanel}
|{\cal M}^s_L|^2
= \frac{\tilde{J}_0\tilde{J}_0^*\tilde{J}_0'\tilde{J}_0'^*}{|s_0|^2},\quad
|{\cal M}^s_T|^2
= \frac{\tilde{J}_{\perp i}\tilde{J}_{\perp k}^*\tilde{J}_{\perp i}'\tilde{J}_{\perp k}'^*}{|s_\perp|^2},\quad
{\cal M}^s_L{\cal M}_T^{s*}
= \frac{\tilde{J}_0\tilde{J}_{\perp i}^*\tilde{J}_0'\tilde{J}_{\perp i}'^*}{s_0s_\perp^*},
\eea
and the interference terms that couple both channels
\bea\label{eq:matrix_ll_tt_ep_mix}
{\cal M}^t_L {\cal M}^{s*}_L = 
\frac{J_0\tilde{J}_0^*J'_0\tilde{J}_0'^{*}}{t_0 s_0^*},\qquad
{\cal M}^t_T {\cal M}^{s*}_T =
\frac{J_{i\perp}\tilde{J}_{k\perp}^*J_{i\perp}'
\tilde{J}_{k\perp}'^*}{t_\perp s_\perp^*},\\
\label{eq:matrix_lt_ep_mix}
{\cal M}^t_L {\cal M}^{s*}_T =
\frac{J_0\tilde{J}_{k\perp}^*J_0'
\tilde{J}_{k\perp}'^*}{t_0 s_\perp^*},\qquad
{\cal M}^t_T {\cal M}^{s*}_L =
\frac{J_{i\perp} \tilde{J}_0^*
J_{i\perp}' \tilde{J}_0^{'*}}{t_\perp s_0 ^*}.
\eea
Equations~\eqref{eq:matrix_ep_t_chanel} and
\eqref{eq:matrix_ep_s_chanel} can be easily averaged over the spins,
but the averaging of the interference
terms~\eqref{eq:matrix_ll_tt_ep_mix} and \eqref{eq:matrix_lt_ep_mix}
is quite cumbersome. Although all of these terms are generally of the
same order of magnitude for relativistic particles, in this work, we
will neglect, for simplicity, the interference terms and keep only the
squares of the scattering and annihilation diagrams. Consider the
$t$-channel first. The spin-averaged current-current couplings are
given by
\bea\label{eq:currents_av}
\frac{1}{2}\sum\limits_{s_1s_3}J_\mu J_\nu^* 
&=& \frac{e^{*2}}{2\ep_1\ep_3}
 \Big[ p_{1\mu} p_{3\nu} + p_{1\nu} p_{3\mu}  - g_{\mu \nu} (\ep_1\ep_3 -\vecp_1\cdot\vecp_3 - m^2 )\Big],\\
\label{eq:currents_av'}
\frac{1}{2}\sum\limits_{s_2s_4}J'_\mu J'^*_\nu 
&=& \frac{e^{*2}}{2\ep_2\ep_4}
 \Big[ p_{2\mu} p_{4\nu} + p_{2\nu} p_{4\mu}  - g_{\mu \nu} (\ep_2\ep_4 -\vecp_2\cdot\vecp_4 - m^2 )\Big].
 \eea
Substituting the relevant components of the currents into Eq.~\eqref{eq:matrix_ep_t_chanel}, 
we obtain for the $t$-channel terms
\bea\label{eq:matrix_L}
\frac{1}{4}\sum\limits_{\rm spins}|{\cal M}^t_L|^2 &=& \frac{e^{*4}}{4\ep_1\ep_2\ep_3\ep_4}
\frac{(\ep_1 \ep_3 + \vecp_1\cdot\vecp_3 + m^2)(\ep_2 \ep_4 + \vecp_2\cdot\vecp_4 + m^2)}{|t_0|^2},\\
\label{eq:matrix_T}
\frac{1}{4}\sum\limits_{\rm spins} |{\cal M}^t_T|^2 &=&
\frac{e^{*4}}{4\ep_1\ep_2\ep_3\ep_4|t_\perp|^2}
\Big[4(\vecp^\perp\cdot\vecp'^\perp)^2+2\left(\ep_1 \ep_3 -\vecp_1\cdot\vecp_3 - m^2 \right)\left(\ep_2 \ep_4 -\vecp_2\cdot\vecp_4 - m^2 \right)
\nonumber\\
&&+2(\vecp^\perp)^2 
\left(\ep_2 \ep_4 -\vecp_2\cdot\vecp_4 - m^2 \right)+ 2(\vecp'^\perp)^2 
\left(\ep_1 \ep_3 -\vecp_1\cdot\vecp_3 - m^2 \right)\Big],\quad\\
\label{eq:matrix_LT}
\frac{1}{4}\sum\limits_{\rm spins}
{\cal M}^t_L{\cal M}_T^{t*}
&=&\frac{e^{*4}}{4\ep_1\ep_2\ep_3\ep_4} 
\frac{(\vecp^\perp\cdot\vecp'^\perp )(\ep_1\ep_2 +\ep_1\ep_4+\ep_2\ep_3+\ep_3\ep_4)}
{t_0t^*_\perp}.
\eea
In Eqs.~\eqref{eq:matrix_T} and \eqref{eq:matrix_LT} we used the fact that $\delta^\perp_{ik}=\delta_{ik}-{q}_i{q}_k/\vecq^2$, and  $\vecp_3^\perp =\vecp_1^\perp \equiv \vecp^\perp$,
$\vecp_4^\perp =\vecp_2^\perp \equiv \vecp'^\perp$,
which follow from the relations $\vecp_3 =\vecp_1-\vecq$, $\vecp_4 =\vecp_2+\vecq$. 

We now modify the numerators using the substitutions $\vecp_3 =
\vecp-\vecq$, $\vecp_4 = \vecp'+\vecq$ and $\ep_3 = \ep-\omega$,
$\ep_4 = \ep'+\omega$ (with $p\equiv p_1$ and  $p'\equiv p_2$)
\bea
&& \ep_1 \ep_3 + \vecp_1\cdot\vecp_3 + m^2 = 2\ep^2-\ep\omega - \vecp \cdot \vecq,\nonumber\\
&& \ep_2 \ep_4 + \vecp_2\cdot\vecp_4 + m^2 = 2\ep'^2+\ep'\omega + \vecp' \cdot \vecq,\nonumber\\
&& \ep_1 \ep_3 - \vecp_1\cdot\vecp_3 - m^2 = \vecp \cdot \vecq - \ep\omega,\nonumber\\
&& \ep_2 \ep_4 - \vecp_2\cdot\vecp_4 - m^2 = \ep'\omega-\vecp' \cdot \vecq,\nonumber\\
&&(\vecp^\perp\cdot\vecp'^\perp )(\ep_1\ep_2 +\ep_1\ep_4+\ep_2\ep_3+\ep_3\ep_4)
=(\vecp^\perp\cdot\vecp'^\perp )
(2\ep-\omega)(2\ep'+\omega).
\nonumber
\eea
Substituting these expressions in  Eqs.~\eqref{eq:matrix_L}--\eqref{eq:matrix_LT}, for the spin-averaged square of the $t$-channel matrix element~\eqref{eq:matrix_t_squared} we  obtain
\bea\label{eq:probability_ep2}
\frac{1}{4}\sum\limits_{\rm spins}|{\cal M}^{ep,t}_{12\to 34}|^2 &=& \frac{e^{*4}}{4\ep\ep'(\ep-\omega)(\ep'+\omega)}\Bigg\{\frac{(2\ep^2-\ep\omega - \vecp \cdot \vecq)(2\ep'^2+\ep'\omega + \vecp' \cdot \vecq)}{|t_0|^2}\nonumber\\
 && -2{\rm Re}\frac{(\vecp^\perp\cdot\vecp'^\perp )(2\ep-\omega)(2\ep'+\omega)}{(t_0t^*_\perp)}+ \frac{1}{|t_\perp|^2}
\Big[2\left(\vecp \cdot \vecq - \ep\omega \right)\left(\ep'\omega-\vecp' \cdot \vecq \right)\nonumber\\
&& +4(\vecp^\perp\cdot\vecp'^\perp)^2+2(\vecp^\perp)^2 
\left(\ep'\omega-\vecp' \cdot \vecq\right)+2(\vecp'^\perp)^2 
\left(\vecp \cdot \vecq - \ep\omega \right)\Big]\Bigg\}.
\eea
\begin{figure}[t]
\includegraphics[width=7cm, keepaspectratio]{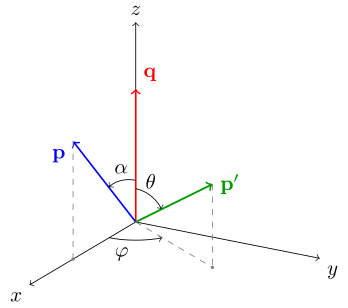}
\caption[] {The relative orientation of three vectors $\vecq$, $\vecp$ and $\vecp'$ with angles defined by Eq.~\eqref{eq:angles_ep1}.
}
\label{fig:angles}
\end{figure}
Defining the three angles between the vectors $\vecq$, $\vecp$, and $\vecp'$ by (see Fig.~\ref{fig:angles})
\bea\label{eq:angles_ep1}
\vecp \cdot \vecq = p q \cos \alpha, 
\quad \vecp'\cdot \vecq = p' q\cos \theta,
\quad \vecp_\perp\cdot \vecp'_\perp = p_\perp p'_\perp \cos\varphi,
\eea
we obtain, for Eq.~\eqref{eq:probability_ep2},
\bea\label{eq:probability_ep3}
\frac{1}{4}\sum\limits_{\rm spins}|{\cal M}^{ep,t}_{12\to 34}|^2 &=& \frac{e^{*4}}{4\ep\ep'(\ep-\omega)(\ep'+\omega)}\Bigg\{\frac{(2\ep^2-\ep\omega - p q \cos \alpha)(2\ep'^2+\ep'\omega + p' q\cos \theta)}{|t_0|^2}\nonumber\\
 && -2{\rm Re}\frac{(p p'\sin\alpha\sin\theta\cos\varphi)(2\ep-\omega)(2\ep'+\omega)}
{(t_0t^*_\perp)}\nonumber\\
&&+ \frac{1}{|t_\perp|^2}\Big[4(p p'\sin\alpha\sin\theta\cos\varphi)^2 +2\left(p q \cos \alpha- \ep\omega \right)\left(\ep'\omega-p' q\cos \theta \right) \nonumber\\
&&
+ 2p^2\sin^2\alpha
\left(\ep'\omega-p' q\cos \theta\right)+2p'^2\sin^2\theta  
\left(p q \cos \alpha - \ep\omega \right)\Big]\Bigg\}.
\eea
The square of the annihilation diagram can be computed in an analogous
manner. Performing the spin averaging of the relevant terms in
Eq.~\eqref{eq:matrix_ep_s_chanel}, substituting
Eq.~\eqref{eq:currents_tilde_ep}, and using the same techniques as
above, we obtain
\bea\label{eq:currents_av_2}
\frac{1}{2}\sum\limits_{s_1s_2}\tilde{J}_\mu \tilde{J}_\nu^* &=& \frac{e^{*2}}{2\ep_1\ep_2}\Big[ p_{1\mu} p_{2\nu} + p_{1\nu} p_{2\mu} - g_{\mu \nu} (\ep_1\ep_2 -\vecp_1\cdot\vecp_2 + m^2 )\Big],\\
\label{eq:currents_av'_2}
\frac{1}{2}\sum\limits_{s_3s_4}\tilde{J}'_\mu \tilde{J}'^*_\nu &=& \frac{e^{*2}}{2\ep_3\ep_4}\Big[ p_{3\mu} p_{4\nu} + p_{3\nu} p_{4\mu} - g_{\mu \nu} (\ep_3\ep_4 -\vecp_3\cdot\vecp_4 + m^2)\Big],
\eea
and
\bea
\frac{1}{4}\sum\limits_{\rm spins}|{\cal M}^{ep,s}_{12\to 34}|^2&=& \frac{e^{*4}}{4\ep_1\ep_2\ep_3\ep_4}\Bigg\{\frac{(\ep_1 \ep_2 + \vecp_1\cdot\vecp_2 - m^2)(\ep_3 \ep_4 + \vecp_3\cdot\vecp_4 - m^2)}{|s_0|^2}\nonumber\\
 && -2 {\rm Re}\frac{(\vecp_1^\perp\cdot\vecp_3^\perp )(\ep_1\ep_3 +\ep_2\ep_4-\ep_1\ep_4-\ep_2\ep_3)}{(s_0s^*_\perp)}
 \nonumber\\
&& +\frac{1}{|s_\perp|^2}
\Big[4(\vecp_1^\perp\cdot\vecp_3^\perp)^2+2\left(\ep_1 \ep_2 -\vecp_1\cdot\vecp_2 + m^2 \right)\left(\ep_3 \ep_4 -\vecp_3\cdot\vecp_4 + m^2 \right)\nonumber\\
&& -2(\vecp_1^\perp)^2 
\left(\ep_3 \ep_4 -\vecp_3\cdot\vecp_4 + m^2 \right) -2(\vecp_3^\perp)^2 
\left(\ep_1 \ep_2 -\vecp_1\cdot\vecp_2 + m^2 \right)\Big]\Bigg\}.
\eea
Defining $\ep''=\ep_3$, $p''=p_3$, $\tilde{\omega}=\ep+\ep_2=\ep''+\ep_4$, and $\tilde{\vecq}=\vecp+\vecp_2=\vecp''+\vecp_4$, and introducing three new angles by
\bea\label{eq:angles_ep2}
\vecp \cdot \tilde{\bm q} = p \tilde{q} \cos \alpha', 
\quad \vecp''\cdot \tilde{\bm q} = p'' \tilde{q}\cos \theta',
\quad \vecp_\perp\cdot \vecp''_\perp = p_\perp p''_\perp \cos\varphi',
\eea
we obtain the final expression for the annihilation amplitude 
\bea
\frac{1}{4}\sum_{\rm spins} |{\cal M}^{ep,s}_{12\to 34}|^2
&=& \frac{e^{*4}}{4 \, \epsilon \, \epsilon'' \, (\tilde{\omega}-\epsilon)(\tilde{\omega}-\epsilon'')}
\Bigg\{ 
\frac{(2\epsilon^2 - \epsilon \tilde{\omega} - p \tilde{q} \cos \alpha')
      (2\epsilon''^2 - \epsilon'' \tilde{\omega} - p'' \tilde{q} \cos \theta')}{|s_0|^2} 
\nonumber\\
&& - 2 \, \mathrm{Re} \frac{(p \, p'' \, \sin \alpha' \, \sin \theta' \, \cos \varphi') 
      (2\epsilon - \tilde{\omega}) (2\epsilon'' - \tilde{\omega})}{s_0 s_\perp^*} 
\nonumber\\
&& + \frac{1}{|s_\perp|^2} \Big[ 
      4 \, (p \, p'' \, \sin \alpha' \, \sin \theta' \, \cos \varphi')^2 
      + 2 \, (p \tilde{q} \cos \alpha' - \epsilon \tilde{\omega}) 
            (p'' \tilde{q} \cos \theta' - \epsilon'' \tilde{\omega}) 
\nonumber\\
&& + 2 \, p^2 \, \sin^2 \alpha' \, (p'' \tilde{q} \cos \theta' - \epsilon'' \tilde{\omega}) 
      + 2 \, p''^2 \, \sin^2 \theta' \, (p \tilde{q} \cos \alpha' - \epsilon \tilde{\omega})
      \Big] 
\Bigg\}.
\eea

We next show that the matrix element for $pe$ collisions is the same
as the one for $ep$ collisions. To check this, it is sufficient to
interchange electrons and positrons, which corresponds to the
following choice of Dirac 4-currents instead of
Eqs.~\eqref{eq:currents_ep} and \eqref{eq:currents_tilde_ep},
\bea
J^{\mu}=-e^*\bar{v}_1 \gamma^\mu v_3,\quad
J'^{\mu}=-e^*\bar{u}_4 \gamma^\mu u_2,\\
\tilde{J}^{\mu}=-e^*\bar{v}_1 \gamma^\mu u_2,\quad
\tilde{J}'^{\mu}=-e^*\bar{u}_4 \gamma^\mu v_3.
\eea
Starting with these currents, a computation analogous to the one above
leads  to exactly the same Eqs.~\eqref{eq:currents_av} and
\eqref{eq:currents_av'}, as well as Eqs.~\eqref{eq:currents_av_2} and
\eqref{eq:currents_av'_2}. Therefore,  the final forms of the
scattering and annihilation terms in the matrix element are the same.

\section{Evaluating the electron-positron collision rate}
\label{app:nu_ep}

Substituting the electron-positron scattering
amplitude~\eqref{eq:matrix_squared} in the collision
rate~\eqref{eq:nu_ep1}, we obtain
\bea\label{eq:nu_ep}
\nu_{ep}&=& \frac{(2\pi)^{-8}}{2l^-}\!\int\! d\bm p_1\, d\bm p_2\, d\bm p_3\, d\bm p_4\, \delta^{(4)}(p_1+p_2-p_3-p_4) \vecq^2\nonumber\\
&\times&
f^{0-}_1f_2^{0+}(1-f^{0-}_3)(1-f_4^{0+}) 
\Big(|{\cal M}^{ep,t}_{12\to 34}|^2+|{\cal M}^{ep,s}_{12\to34}|^2\Big),
\eea
where we dropped the terms that mix the $t$ and $s$-channels. 
After introducing dummy integration over transferred momentum and energy for each channel separately, we write
$\nu_{ep}=\nu^t_{ep}+\nu^s_{ep}$, where $\nu^t_{ep}$ and $\nu^s_{ep}$
denote the scattering and annihilation parts of the integrals,
respectively; they  are given by
\bea\label{eq:nu_ep_t}
\nu^t_{ep}  
&=& \frac{(2\pi)^{-8}}{2l^-}\!\int\! d\omega\, d\bm q\!\int\! d\bm p\, d\bm p'\,
\delta(\ep-\ep_3-\omega)\delta(\ep'-\ep_4+\omega)\nonumber\\
&& \times \vecq^2 f^{0-}_1f_2^{0+}(1-f^{0-}_3)(1-f_4^{0+}) |{\cal M}^{ep,t}_{12\to 34}|^2,\\
\label{eq:nu_ep_s}
\nu^s_{ep} 
&=& \frac{(2\pi)^{-8}}{2l^-}\!\int\! d\tilde{\omega}\, d\tilde{\bm q}\!\int\! d\bm p\, d\bm p''\,\delta(\ep+\ep_2-\tilde{\omega})\delta(\ep''+\ep_4-\tilde{\omega})\nonumber\\
&&\times \vecq^2 f^{0-}_1f_2^{0+}(1-f^{0-}_3)(1-f_4^{0+}) |{\cal M}^{ep,s}_{12\to 34}|^2.
\eea
Note that in Eq.~\eqref{eq:nu_ep_t}, the energies $\ep_3$ and $\ep_4$ should be calculated for the momentum values $\vecp_3=\vecp-\vecq$ and $\vecp_4=\vecp'+\vecq$, respectively, and in Eq.~\eqref{eq:nu_ep_s} $\ep_2$ and $\ep_4$ should be calculated for $\vecp_2=\tilde{\vecq}-\vecp$ and $\vecp_4=\tilde{\vecq}-\vecp''$. Here, we renamed $\bm p_1=\bm p$, $\bm p_2=\bm p'$, $\bm p_3=\bm p''$, and $\ep_1=\ep$, $\ep_2=\ep'$, $\ep_3=\ep''$. The four $\delta$-functions in Eqs.~\eqref{eq:nu_ep_t} and \eqref{eq:nu_ep_s} can be written as
\bea\label{eq:delta_ep1}
\delta(\ep-\ep_3-\omega)
&=& \delta\left(\ep-\omega-\sqrt{\ep^2+q^2-2pq\cos\alpha}\right)
=\frac{\ep-\omega}{pq}
\delta(\cos\alpha-x_0)
\theta(\ep-\omega),\\
\label{eq:delta_ep2}
\delta(\ep'-\ep_4+\omega)
&=& \delta\left(\ep'+\omega-\sqrt{\ep'^2+q^2+2p'q\cos\theta}\right)
=\frac{\ep'+\omega}{p'q}\delta(\cos\theta-y_0)\theta(\ep'+\omega),\\
\label{eq:delta_ep1_an}
\delta(\ep+\ep_2-\tilde{\omega})
&=& \delta\left(\ep-\tilde{\omega}+\sqrt{\ep^2+\tilde{q}^2-2p\tilde{q}\cos\alpha'}\right)
= \frac{\tilde{\omega}-\ep}{p\tilde{q}}
\delta(\cos\alpha'-\tilde{x}_0)
\theta(\tilde{\omega}-\ep),\\
\label{eq:delta_ep2_an}
\delta(\ep''+\ep_4-\tilde{\omega})
&=& \delta\left(\ep''-\tilde{\omega}+\sqrt{\ep''^2+\tilde{q}^2-2p''\tilde{q}\cos\theta'}\right)
= \frac{\tilde{\omega}-\ep''}{p''\tilde{q}}
\delta(\cos\theta'-\tilde{y}_0)
\theta(\tilde{\omega}-\ep''),\qquad
\eea
where we used the definitions of angles~\eqref{eq:angles_ep1} and \eqref{eq:angles_ep2} and defined 
\bea\label{eq:xy_0}
x_0=\frac{q^2-\omega^2+2\ep\omega}{2pq}, \qquad  y_0=\frac{\omega^2-q^2+2\ep'\omega}{2p'q},\\
\label{eq:xy_0_tilde}
\tilde{x}_0=\frac{\tilde{q}^2-\tilde{\omega}^2+2\ep\tilde{\omega}}{2p\tilde{q}},
\qquad  
\tilde{y}_0=\frac{\tilde{q}^2-\tilde{\omega}^2+2\ep''\tilde{\omega}}{2p''\tilde{q}}.
\eea
Using these relations and substituting the matrix elements from Eqs.~\eqref{eq:probability_ep4_t} and \eqref{eq:probability_ep4_s} into Eqs.~\eqref{eq:nu_ep_t} and \eqref{eq:nu_ep_s}, we obtain 
\bea\label{eq:nu_ep3}
\nu^t_{ep} &=& \frac{(2\pi)^{-8}}{2l^-}4\pi\!\int_m^\infty\! \ep d \ep\! \int_m^\infty\! \ep' d \ep'\!\int_{-\ep'}^{\ep} d\omega\, f^{0-}_1f_2^{0+}(1-f^{0-}_3)(1-f_4^{0+})\nonumber\\
&& \times 2\pi\!\int_0^{\infty}\!
dq \!\int_{\pi}^{0}{d(\cos\alpha)} \delta(\cos\alpha-x_0)I_\Omega,\\
\label{eq:nu_ep3_s}
\nu^s_{ep} &=& \frac{(2\pi)^{-8}}{2l^-}4\pi\!\int_m^\infty\! \ep d \ep\! \int_m^\infty\! \ep'' d \ep''\!\int_{\omega_0}^{\infty} d\tilde{\omega}\, f^{0-}_1f_2^{0+}(1-f^{0-}_3)(1-f_4^{0+})\nonumber\\
&& \times 2\pi\!\int_0^{\infty}\!
d\tilde{q} \!\int_{\pi}^{0}{d(\cos\alpha')} \delta(\cos\alpha'-\tilde{x}_0)\tilde{I}_\Omega,
\eea
where we used the relation $pdp=\ep d\ep$ and defined $\omega_0 = {\rm
  max}\{\ep,\ep''\}$. We also introduced  the angular integrals as
\bea\label{eq:angle_integral}
I_{\Omega}&=&\frac{e^{*4}}{4\ep\ep'}\!\int_{\pi}^0 d (\cos\theta)  \delta(\cos\theta-y_0)\! \int_0^{2\pi}\! d\varphi
 \Biggl[\frac{C_0(\alpha, \theta)}{|t_0|^2}
+\frac{C_2(\alpha,\theta)+4C_1^2(\alpha,\theta)\cos^2\varphi}{|t_\perp|^2}\nonumber\\
&& -2{\rm Re}\frac{(2\ep-\omega)(2\ep'+\omega)C_1(\alpha,\theta)\cos\varphi}{t_0t^*_\perp}
\Biggr]q^2,\\
\label{eq:angle_integral_an}
\tilde{I}_{\Omega}&=&\frac{e^{*4}}{4\ep\ep''}\!\int_{\pi}^0 d
(\cos\theta')  \delta(\cos\theta'-\tilde{y}_0)\! \int_0^{2\pi}\! d\varphi
 \Biggl[\frac{\tilde{C}_0(\alpha', \theta')}{|s_0|^2}
 +\frac{\tilde{C}_2(\alpha',\theta')
   +4\tilde{C}_1^2(\alpha',\theta')\cos^2\varphi'}{|s_\perp|^2}\nonumber\\
&& -2{\rm Re}\frac{(2\ep-\tilde{\omega})(2\ep''-\tilde{\omega})\tilde{C}_1(\alpha',\theta')\cos\varphi'}{s_0s^*_\perp}
\Biggr]q^2.
\eea
We next carry out the azimuthal ($\varphi,\varphi'$)-integrations in
Eqs.~\eqref{eq:angle_integral} and \eqref{eq:angle_integral_an} after
substituting 
\bea
{\vecq}^2 = (\vecp - \vecp'')^2= p^2+p''^2 -2p p''(\cos\alpha\cos\theta+\sin\alpha\sin\theta\cos\varphi),
\eea
where we used Eq.~\eqref{eq:angles_ep1}. A similar expression
follows from Eq.~\eqref{eq:angles_ep2} for the primed
angles.
The integrals of $\cos\varphi$ and $\cos^3\varphi$ over the interval
$[0,2\pi]$ vanish, while the integral of $\cos^2\varphi$ gives $\pi$
and the terms that do not depend on $\varphi$ integrate to
$2\pi$. Therefore,
\bea\label{eq:angle_integral1}
I_{\Omega} &=& 2\pi \frac{e^{*4}}{4\ep\ep'}\! \int_{\pi}^0 d (\cos\theta)  \delta(\cos\theta-y_0)
\Biggl[
 \frac{C_0(\alpha, \theta)}{|t_0|^2}+\frac{C_2(\alpha,\theta)+2C_1^2(\alpha,\theta)}{|t_\perp|^2}\Biggr]q^2,\\
\label{eq:angle_integral1_an}
\tilde{I}_{\Omega} &=& 2\pi \frac{e^{*4}}{4\ep\ep''}\! \int_{\pi}^0 d (\cos\theta)  \delta(\cos\theta-\tilde{y}_0)
\Bigg\{\Biggl[
 \frac{\tilde{C}_0(\alpha, \theta)}{|s_0|^2}+\frac{\tilde{C}_2(\alpha,\theta)+2\tilde{C}_1^2(\alpha,\theta)}{|s_\perp|^2}\Biggr]\nonumber\\
&\times&\big(p^2+p''^2-2pp''\cos\alpha\cos\theta\big)+2{\rm Re}\frac{pp''(2\ep-\tilde{\omega})(2\ep''-\tilde{\omega})\tilde{C}_1(\alpha,\theta)\sin\alpha\sin\theta}{s_0s^*_\perp}\Bigg\},\qquad
\eea
where we dropped primes on angles in the second expression, and
\bea\label{eq:angle_integral2}
\int_{\pi}^{0}{d(\cos\alpha)} \delta(\cos\alpha-x_0)I_\Omega
 &=& 2\pi \frac{e^{*4}}{4\ep\ep'} 
\Biggl[
 \frac{{D}_0}{|t_0|^2}+\frac{{D}_2+2p^2p'^2(1-x_0^2)(1-y_0^2)}{|t_\perp|^2}\Biggr]q^2
 \theta(1-\vert x_0\vert)\theta(1-\vert y_0\vert),\\
\label{eq:angle_integral2_an}
\int_{\pi}^{0}{d(\cos\alpha)} \delta(\cos\alpha-\tilde{x}_0)\tilde{I}_\Omega
 &=& 2\pi \frac{e^{*4}}{4\ep\ep''} 
\Bigg\{\Biggl[
 \frac{{\tilde{D}}_0}{|s_0|^2}+\frac{{\tilde{D}}_2+2p^2p''^2(1-\tilde{x}_0^2)(1-\tilde{y}_0^2)}{|s_\perp|^2}\Biggr]\big(p^2+p''^2-2pp''\tilde{x}_0\tilde{y}_0\big) \nonumber\\
&+&2{\rm Re}\frac{p^2p''^2(2\ep-\tilde{\omega})(2\ep''-\tilde{\omega})(1-\tilde{x}_0^2)(1-\tilde{y}_0^2)}{s_0s^*_\perp}\Bigg\}\theta(1-\vert \tilde{x}_0\vert)\theta(1-\vert\tilde{y}_0\vert).\qquad\qquad
\eea
Here we used Eqs.~\eqref{eq:C0}--\eqref{eq:C2_tilde} and
defined new functions by
\bea\label{eq:D0_new}
{D}_0 &=& 
(2\ep^2-\ep\omega - p q x_0)(2\ep'^2+\ep'\omega + p'qy_0),\\
\label{eq:D2_new}
{D}_2 &=&
2p^2(1-x_0^2)
\left(\ep'\omega-p' qy_0\right)+2p'^2(1-y_0^2)  
\left(p q x_0 - \ep\omega \right)\nonumber\\
&& +2\left(p q x_0- \ep\omega \right)\left(\ep'\omega-p'qy_0\right),\\
\label{eq:D0_new_tilde}
{\tilde{D}}_0 &=& 
(2\ep^2-\ep\tilde{\omega} - p \tilde{q} \tilde{x}_0)(2\ep''^2-\ep''\tilde{\omega} - p'' \tilde{q} \tilde{y}_0),\\
\label{eq:D2_new_tilde}
{\tilde{D}}_2 &=&
2p^2(1-\tilde{x}_0^2)
\left(p'' \tilde{q}\tilde{y}_0-\ep''\tilde{\omega}\right)+2p''^2(1-\tilde{y}_0^2)  
\left(p \tilde{q} \tilde{x}_0 - \ep\tilde{\omega} \right)\nonumber\\
&& +2\left( \ep\tilde{\omega}-p \tilde{q} \tilde{x}_0 \right)\left(\ep''\tilde{\omega}-p'' \tilde{q}\tilde{y}_0\right).
\eea
Substituting Eqs.~\eqref{eq:angle_integral2} and \eqref{eq:angle_integral2_an} back into Eqs.~\eqref{eq:nu_ep3} and \eqref{eq:nu_ep3_s}, we obtain
\bea\label{eq:nu_ep4}
\nu^t_{ep} &=& (2\pi)^{-5}\frac{e^{*4}}{4l^-}\!\int_m^\infty\! d\ep\! \int_m^\infty\! d\ep'\! \int_{-\ep'}^{\ep} d\omega\, f^{0-}_1f_2^{0+}(1-f^{0-}_3)(1-f_4^{0+}) \int_0^\infty\! dq\nonumber\\
&& \times \Biggl[
 \frac{{D}_0}{|t_0|^2}+\frac{{D}_2+2p^2p'^2(1-x_0^2)(1-y_0^2)}{|t_\perp|^2}\Biggr]q^2\theta(1-\vert x_0\vert)\theta(1-\vert y_0\vert),\\
\label{eq:nu_ep4_an} 
\nu^s_{ep} &=& (2\pi)^{-5}\frac{e^{*4}}{4l^-}\!\int_m^\infty\! d\ep\! \int_m^\infty\! d\ep''\! \int_{\omega_0}^{\infty} d\tilde{\omega}\, f^{0-}_1f_2^{0+}(1-f^{0-}_3)(1-f_4^{0+}) \int_0^\infty\! d\tilde{q}\nonumber\\
&& \times \Bigg\{\Biggl[
 \frac{{\tilde{D}}_0}{|s_0|^2}+\frac{{\tilde{D}}_2+2p^2p''^2(1-\tilde{x}_0^2)(1-\tilde{y}_0^2)}{|s_\perp|^2}\Biggr]\big(p^2+p''^2-2pp''\tilde{x}_0\tilde{y}_0\big)\nonumber\\
 &&+2{\rm Re}\frac{p^2p''^2(2\ep-\tilde{\omega})(2\ep''-\tilde{\omega})(1-\tilde{x}_0^2)(1-\tilde{y}_0^2)}{s_0s^*_\perp}\Bigg\}\theta(1-\vert \tilde{x}_0\vert)\theta(1-\vert\tilde{y}_0\vert).
 \eea

The $\theta$-functions in Eq.~\eqref{eq:nu_ep4} give two maxima and two minima for $q$:
$q_-\leq q\leq q_+$, $q'_-\leq q\leq q'_+$, where
\bea
q_{\pm} = \left\vert  \sqrt{(\ep-\omega)^2-m^2} \pm \sqrt{\ep^2-m^2}\right\vert,
\qquad q'_{\pm} = \left\vert  \sqrt{(\ep'+\omega)^2-m^2} \pm \sqrt{\ep'^2-m^2}\right\vert.
\eea
Similarly, the $\theta$-functions in Eq.~\eqref{eq:nu_ep4_an} imply $\tilde{q}_-\leq \tilde{q}\leq \tilde{q}_+$, $\tilde{q}'_-\leq \tilde{q}\leq \tilde{q}'_+$, with
\bea
\tilde{q}_{\pm} = \left\vert  \sqrt{(\ep-\tilde{\omega})^2-m^2} \pm   \sqrt{\ep^2-m^2}\right\vert,
\qquad \tilde{q}'_{\pm} = \left\vert  \sqrt{(\ep''-\tilde{\omega})^2-m^2} \pm   \sqrt{\ep''^2-m^2}\right\vert.
\eea
Therefore, the final upper and lower limits of the $q$- and $\tilde{q}$-integrations will be given, respectively, by $Q_-={\rm max}(q_-, q'_-)$, $Q_+={\rm min}(q_+, q'_+)$, and $\tilde{Q}_-={\rm max}(\tilde{q}_-, \tilde{q}'_-)$, $\tilde{Q}_+={\rm min}(\tilde{q}_+, \tilde{q}'_+)$.
Note also, that, to have real values for $q$, we need additional conditions $\omega \le \ep-m$ and $\omega \ge -\ep'+m$, and 
to have real values for $\tilde{q}$ we need the conditions $\tilde{\omega} \ge \ep+m$ and $\tilde{\omega} \ge \ep''+m$.
Implementing these limits and also substituting the expressions~\eqref{eq:D0_new}--\eqref{eq:D2_new_tilde} in Eqs.~\eqref{eq:nu_ep4} and \eqref{eq:nu_ep4_an}, we obtain
\bea\label{eq:nu_ep6}
\nu_{ep}^t &=& (2\pi)^{-5}\frac{e^{*4}}{4l^-}\!\int_m^\infty\! d\ep\! \int_m^\infty\! d\ep'\!\int_{m-\ep'}^{\ep-m} d\omega\,f^{0-}_1f_2^{0+}(1-f^{0-}_3)(1-f_4^{0+}) \int_{Q_-}^{Q_+}  dq \nonumber\\
&\times& \Bigg[\frac{\big[(2\ep-\omega)^2-q^2\big]\big[(2\ep'+\omega)^2-q^2\big]}{4|t_0|^2}+\frac{1}{|t_\perp|^2}\Big(p^2(1-x_0^2)(q^2-\omega^2)\nonumber\\
&+&p'^2(1-y_0^2)(q^2-\omega^2) +\frac{1}{2}(q^2-\omega^2)^2+2p^2p'^2(1-x_0^2)(1-y_0^2)\Big)\Bigg]q^2,\nonumber\\
\label{eq:nu_ep_an}
\nu_{ep}^s &=& (2\pi)^{-5}\frac{e^{*4}}{4l^-}\!\int_m^\infty\! d\ep\! \int_m^\infty\! d\ep''\!\int_{\omega_0+m}^{\infty} d\tilde{\omega}\,f^{0-}_1f_2^{0+}(1-f^{0-}_3)(1-f_4^{0+}) \int_{\tilde{Q}_-}^{\tilde{Q}_+}  d\tilde{q} \nonumber\\
&\times& \Bigg\{\Bigg[\frac{\big[(2\ep-\tilde{\omega})^2-\tilde{q}^2\big]\big[(2\ep''-\tilde{\omega})^2-\tilde{q}^2\big]}{4|s_0|^2}+\frac{1}{|s_\perp|^2}\Big(p^2(1-\tilde{x}_0^2)(\tilde{q}^2-\tilde{\omega}^2)\nonumber\\
&+&p''^2(1-\tilde{y}_0^2)(\tilde{q}^2-\tilde{\omega}^2) +\frac{1}{2}(\tilde{q}^2-\tilde{\omega}^2)^2+2p^2 p''^2(1-\tilde{x}_0^2)(1-\tilde{y}_0^2)\Big)\Bigg]\nonumber\\
&\times&\big(p^2+p''^2-2pp''\tilde{x}_0\tilde{y}_0\big)+2{\rm Re}\frac{p^2p''^2(2\ep-\tilde{\omega})(2\ep''-\tilde{\omega})(1-\tilde{x}_0^2)(1-\tilde{y}_0^2)}{s_0s^*_\perp}\Bigg\},
\eea
where we used Eqs.~\eqref{eq:xy_0} and \eqref{eq:xy_0_tilde}.
Substituting $e^{*2}=4\pi e^2$, we obtain the final expression~\eqref{eq:nu_ep_final} for the electron-positron collision rate. 

\end{widetext}

\providecommand{\href}[2]{#2}\begingroup\raggedright\endgroup

\end{document}